\documentclass{arxivpaper}

\usepackage{multirow}
\usepackage{bm}
\usepackage{braket}

\definecolor{colgreen}{HTML}{F4F8F1}
\definecolor{colcream}{HTML}{FDFAEF}
\definecolor{colmauve}{HTML}{FAF5F7}
\definecolor{colsteel}{HTML}{F4F8FB}
\definecolor{colpeach}{HTML}{FDF4EF}
\definecolor{colteal}{HTML}{EFF8F7}
\definecolor{hdrgreen}{HTML}{CDDEC4}
\definecolor{hdrcream}{HTML}{F3E4B6}
\definecolor{hdrmauve}{HTML}{E1CAD4}
\definecolor{hdrsteel}{HTML}{C6D8E6}
\definecolor{hdrpeach}{HTML}{EDCDB8} 
\definecolor{hdrteal}{HTML}{BCDDD8} 
\definecolor{rowhilite}{HTML}{EDEDF8}

\title{Q-Capsule: A Localized Capsule-Based Quantum Neural Architecture for Barren Plateau Mitigation}

\author{%
  \parbox{\dimexpr\textwidth-2\tabcolsep\relax}{\centering
    Awal Ahmed Fime\textsuperscript{1,*},
    Tasfia Zaman Samiha\textsuperscript{2,*},
    Saika Zaman\textsuperscript{1},\\[3pt]
    Dimitris Pados\textsuperscript{1},
    George Sklivanitis\textsuperscript{1},
    Abdur R. Shahid\textsuperscript{3},
    Ahmed Imteaj\textsuperscript{1,\ensuremath{\dagger}}\\[9pt]
    {\small\textsuperscript{1}Florida Atlantic University, Boca Raton, Florida, USA\\
    \textsuperscript{2}Khulna University of Engineering \& Technology, Khulna, Bangladesh\\
    \textsuperscript{3}Southern Illinois University Carbondale, Carbondale, Illinois, USA}\\[7pt]
    {\small\textsuperscript{*}Both authors contributed equally to this research.\\
    \textsuperscript{\ensuremath{\dagger}}Corresponding author: \href{mailto:aimteaj@fau.edu}{\texttt{aimteaj@fau.edu}}}
  }%
}
\date{}
\hypersetup{
  pdftitle={Q-Capsule: A Localized Capsule-Based Quantum Neural Architecture for Barren Plateau Mitigation},
  pdfauthor={Awal Ahmed Fime, Tasfia Zaman Samiha, Saika Zaman, Dimitris Pados, George Sklivanitis, Abdur R. Shahid, Ahmed Imteaj}
}

\begin{document}

\begin{abstract}
Variational quantum algorithms are often limited by barren plateaus: gradients vanish as circuit size and depth increase, making quantum neural networks difficult to train. We propose \textsc{Q-Capsule}, a localized capsule-based quantum neural architecture that mitigates this problem through register partitioning, local readout, sparse inter-capsule coupling, trainable data re-uploading, and Quantum Fisher Information Matrix (QFIM)-guided adaptive depth growth. By restricting the dominant support of each observable to a small capsule and controlling inter-capsule entanglement, \textsc{Q-Capsule} preserves useful gradient signals while retaining communication between local quantum representations. As the register width increases, \textsc{Q-Capsule} consistently maintains stable gradient variance, whereas globally entangling baselines exhibit exponential suppression with a log-gradient-variance slope near $-\ln 2$ per qubit. \textsc{Q-Capsule} also produces more structured optimization landscapes, higher parameter efficiency, improved robustness to depolarizing noise, and lower measurement requirements. Its adaptive policy achieves $98.1\%$ accuracy on binary classification and $97.7\%$ on four-class classification, while using approximately $73\%$ fewer two-qubit gates than the fixed-deep model on the multiclass task.
\end{abstract}

\maketitle

\section{Introduction}
\label{sec:introduction}
Variational Quantum Algorithms (VQAs) have emerged as a promising computational paradigm for near-term quantum devices. Their appeal lies in their hybrid quantum–classical structure, where a classical optimizer updates a set of trainable circuit parameters, while a quantum processor evaluates the corresponding objective function through repeated circuit execution and measurement \cite{mcclean2016theory, endo2021hybrid, ramouthar2023hybrid, ajagekar2024variational}. This iterative feedback loop allows VQAs to exploit the computational capabilities of quantum hardware while relying on classical optimization to guide the search toward high-quality solutions. As a result, VQAs are widely regarded as a practical approach for exploring quantum advantage on noisy intermediate-scale quantum devices \cite{bharti2022noisy, moll2018quantum, wu2024qvae}. However, increasing the size, depth, and expressibility of these circuits often makes their optimization increasingly difficult. A central cause of this difficulty is the \emph{barren plateau} phenomenon \cite{larocca2025barren, zhang2022escaping, yao2025avoiding, sack2022avoiding}, in which gradients of the objective function collapse toward zero as the number of qubits increases. When this occurs, gradient-based optimization requires increasingly precise measurements to distinguish useful descent directions from statistical noise, limiting the practical scalability of variational quantum models~\cite{mcclean2018barren}.

The emergence of barren plateaus depends strongly on circuit architecture \cite{holmes2022connecting, pesah2021absence, cervero2023barren},
observable locality \cite{cerezo2021cost, uvarov2021barren, garcia2023barren}, circuit depth \cite{ragone2024lie, mcclean2018barren, cerezo2021cost}, and entanglement structure \cite{marrero2021entanglement, patti2021entanglement, ragone2024lie}. Global cost functions can exhibit exponentially vanishing gradients even for relatively shallow circuits, whereas appropriately chosen local observables can provide more favorable gradient scaling~\cite{sharma2022trainability, cerezo2021cost}. At the same time, highly expressive circuits that approach random state ensembles can suffer from severe gradient suppression~\cite{ holmes2022connecting,sim2019expressibility}. Excessive entanglement can further drive local subsystems toward highly mixed states and reduce the information available to local gradients~\cite{marrero2021entanglement, wiersema2023measurement}, while hardware noise introduces an additional source of depth-dependent gradient decay~\cite{wang2021noise, stilck2021limitations}. These observations suggest that trainability is determined not only by the number of parameters in a quantum model, but also by how those parameters, entangling operations, and measurements are distributed throughout the circuit.

Existing mitigation strategies fall into three broad families. Initialization strategies place the circuit in a near-identity region of parameter space so that training begins away from the plateau~\cite{grant2019initialization, liu2021mitigating, kashif2024alleviating}, and layer-wise schedules defer the optimization of deep, highly expressive blocks~\cite{skolik2021layerwise, allayerwise, sharma2024addressing}; both act on \emph{where} optimization starts rather than on the structure that produces the plateau, and the landscape they eventually enter is unchanged. Structured ansätze instead constrain the circuit itself: quantum convolutional networks achieve at worst polynomially vanishing gradients through hierarchical convolution and pooling~\cite{cong2019quantum,pesah2021absence}, tensor-network-inspired circuits obtain favorable gradient scaling from tree-like or multiscale connectivity~\cite{cervero2023barren, miao2024isometric, barthel2023absence}, and residual quantum networks introduce skip connections that ease gradient
flow across depth without restricting observable
support~\cite{kashif2024resqnets, yi2026q, liu2024laziness}. Finally, cost-function engineering exploits the locality result directly, replacing global observables with local ones~\cite{cerezo2021cost,sharma2022trainability}. Each strategy fixes
its structural commitment, before training begins and holds it constant.

This leaves several critical gaps. First, applying a uniform circuit depth across the entire register assumes that all feature groups require the same representational capacity. In practice, some feature subsets may saturate at shallow depths, while others remain under-parameterized, making a fixed-depth design inefficient. Second, although prior work shows that circuit structure strongly affects trainability, an important trade-off remains unresolved. Highly global circuits can capture non-local correlations but are more susceptible to gradient concentration, whereas fully isolated local circuits can preserve stronger gradients at the cost of limited information exchange across different regions of the quantum state. These limitations motivate architectures that can adapt representational capacity across feature groups while retaining controlled inter-group communication.

To address these limitations, we propose \textsc{Q-Capsule}, a localized capsule-based quantum neural architecture designed around the principle that \emph{entanglement should be treated as a controlled computational resource}. Inspired by the locality and modular organization of classical capsule networks~\cite{sabour2017dynamic}, \textsc{Q-Capsule} partitions an $N$-qubit register into fixed-width quantum capsules. Each capsule contains its own parameters, intra-capsule entangling operations, and local measurement. For a fixed capsule size $c_{\mathrm{size}}$, restricting both the observable and its dominant causal support to a capsule gives the gradient scaling $\operatorname{Var}[\partial_{\theta}\langle O_c\rangle] \in\Omega(2^{-c_{\mathrm{size}}})$, which is independent of the total register width $N$. This design therefore prevents the support of every local objective from automatically growing with the complete Hilbert space as additional qubits are introduced. The architecture nevertheless avoids treating capsules as completely independent components by introducing sparse inter-capsule coupling. Each capsule first gathers and processes information locally, preserving localized trainability, before periodically communicating with neighboring capsules. This controlled interaction occurs only at selected layers rather than at every layer, providing a tunable balance between local processing and global information propagation.

The second key aspect of \textsc{Q-Capsule} is that representational capacity is allocated adaptively rather than through a fixed uniform depth. Each capsule is assigned an independent depth that can increase during training. Growth is controlled using the Quantum Fisher Information Matrix (QFIM), which describes the local geometry of the parameterized quantum state and can reveal effective and redundant optimization directions~\cite{haug2021capacity,meyer2021fisher,liu2019quantum,larocca2022diagnosing}. We derive a per-capsule expressibility score from the effective dimension and conditioning of the QFIM spectrum and add a new layer only when this score begins to saturate. As a result, additional parameters are introduced according to the geometric state of each capsule rather than according to a predetermined global schedule. The architecture further combines this adaptive mechanism with trainable data re-uploading~\cite{perez2020data, easom2021depth, wang2025predictive} and structured initialization~\cite{grant2019initialization,volkoff2021large}, allowing additional depth to increase data-dependent representational capacity while avoiding an unnecessarily difficult initial optimization. 
The resulting architecture provides a middle ground between globally entangling circuits and completely independent local models. Local capsule measurements preserve resolvable gradients, sparse boundary interactions permit information exchange without continuously spreading correlations across the entire register, and QFIM-guided growth expands capacity only when the existing representation approaches saturation. Importantly, these mechanisms address different aspects of trainability: locality controls gradient concentration, sparse coupling controls the growth of inter-capsule entanglement, data re-uploading maintains a strong input-dependent representation, and adaptive growth controls how model capacity is allocated.

Our main contributions are summarized as follows:
\begin{itemize}
    \item We introduce \textsc{Q-Capsule}, a modular quantum neural architecture that partitions the quantum register into fixed-width capsules with independent parameters and local readouts. This construction yields a per-capsule gradient bound governed by $c_{\mathrm{size}}$ rather than by the total register width $N$, providing a scalable architectural mechanism for mitigating width-induced barren plateaus.

    \item We implemented sparse inter-capsule coupling as an \emph{entanglement budget}. Periodic boundary interactions permit information sharing between neighboring capsules while limiting the propagation of global entanglement. Empirically, sparse coupling retains most of the measured inter-capsule quantum mutual information while using substantially fewer coupling gates.

    \item We develop a saturation-gated adaptive growth strategy based on the QFIM. Each capsule can increase its depth independently when its QFIM-derived expressibility score saturates, allowing representational capacity to be allocated according to local optimization geometry rather than through a fixed-depth design.

    \item We provide a comprehensive empirical analysis of trainability using gradient-variance scaling, gradient distributions, cost-landscape geometry, parameter efficiency, component ablations, depolarizing noise, measurement cost, and classification performance. 

    \item We further demonstrate that the observed gradient behavior is not specific to a particular implementation by reproducing the \textsc{Q-Capsule} gradient statistics independently in PennyLane, Qiskit, and Cirq. On binary and four-class digit classification, the adaptive policy also remains competitive across tasks of different complexity without requiring the circuit depth to be selected in advance.
\end{itemize}

The remainder of this paper first reviews related work (Section \ref{sec:related_work}) on structured quantum neural architectures and barren-plateau mitigation. We then present the \textsc{Q-Capsule} architecture (Section \ref{sec:methodology}), including capsule partitioning, local readout, QFIM-guided adaptive growth, sparse inter-capsule coupling, and trainable data re-uploading. Section \ref{sec:results} reports its gradient scaling, optimization geometry, component contributions, resource characteristics, robustness to noise, classification performance, and consistency across multiple quantum software frameworks. Finally, Section \ref{sec:conclusion} concludes the paper by summarizing the main findings and outlining directions for future research.

\section{Related Work}
\label{sec:related_work}

\subsection{Quantum Neural Networks and Structured Architectures}

Variational Quantum Algorithms combine parameterized quantum circuits with classical optimization and form a central paradigm for near-term quantum computing~\cite{cerezo2021variational, qi2024variational, zhao2020measurement}. Quantum Neural Networks (QNNs) apply this framework to tasks such as classification \cite{hafeez2024h, hur2022quantum}, generative modeling \cite{rousselot2024generative, zhang2025denoising}, and quantum chemistry~\cite{roh2024hybrid, smaldone2024quantum}. Although hardware-efficient ansatzes enable implementation on noisy devices~\cite{kandala2017hardware}, their scalability and trainability remain limited. Data re-uploading improves representational capacity by repeatedly encoding classical features without substantially increasing the number of qubits~\cite{aminpour2024strategic, schetakis2025quantum, perez2020data}. More expressive and strongly entangling circuits can approximate Haar-random state distributions~\cite{sim2019expressibility}, but often exhibit smaller gradients and barren plateaus~\cite{holmes2022connecting}. QNN capacity has also been characterized through the Quantum Fisher Information Matrix and effective dimension~\cite{abbas2021power}. Structured alternatives, such as Quantum Convolutional Neural Networks, use hierarchical convolution and pooling to compress quantum information, although fixed pooling schedules may limit adaptation to inputs of varying complexity~\cite{cong2019quantum,rajesh2021quantum, sunil2025quantum}.

\subsection{Barren Plateaus in Variational Quantum Circuits}

The barren-plateau phenomenon was initially characterized through concentration-of-measure arguments showing that sufficiently expressive parameterized quantum circuits can exhibit exponentially suppressed gradient variance as the Hilbert-space dimension grows~\cite{mcclean2018barren}. Subsequent studies have substantially refined this characterization by identifying conditions under which trainability can either deteriorate or be preserved. In particular, theoretical analyses have connected gradient concentration to the choice of objective function~\cite{cerezo2021cost,sharma2022trainability}, circuit architecture~\cite{pesah2021absence,cervero2023barren}, entanglement structure~\cite{marrero2021entanglement}, and the effective expressibility of the ansatz~\cite{holmes2022connecting}. Noise-induced barren plateaus further demonstrate that trainability degradation can arise even when the noiseless circuit itself has favorable optimization properties~\cite{wang2021noise,stilck2021limitations}. More recent Lie-algebraic analyses provide a complementary perspective by relating gradient scaling to the dimension and structure of the dynamical Lie algebra generated by the circuit~\cite{ragone2024lie}.

These theoretical results have motivated a variety of architectural and optimization approaches for improving trainability. Hierarchical circuits such as quantum convolutional neural networks exploit restricted causal structure to obtain favorable gradient scaling under appropriate conditions~\cite{pesah2021absence}. The tensor-network-inspired constructions constrain connectivity and accessible correlations through tree-like or multiscale circuit organizations~\cite{cervero2023barren}. Residual quantum architectures instead introduce alternative information-propagation paths across circuit depth~\cite{kashif2024resqnets}. Other studies modify the optimization trajectory rather than the circuit topology, for example through identity-based initialization~\cite{grant2019initialization} or incremental layer-wise optimization~\cite{skolik2021layerwise}. Collectively, these works demonstrate that controlling circuit structure and optimization complexity can substantially improve gradient behavior. However, the circuit organization and capacity allocation are generally selected before optimization or adjusted according to a predetermined training schedule.

 Our work bridges this gap in barren-plateau mitigation by proposing a capsule-based measurement framework. Instead of strictly choosing between trainability-destroying global observables or representationally limited independent local measurements, our architecture groups operations into localized, structured capsules. Combined with optimized, alternating entangling layers, this balances expressibility and trainability, providing an explicit defense against barren plateau emergence while maintaining high-fidelity objective matching.

\begin{figure}[htb!]
    \centering
    \includegraphics[width=\textwidth]{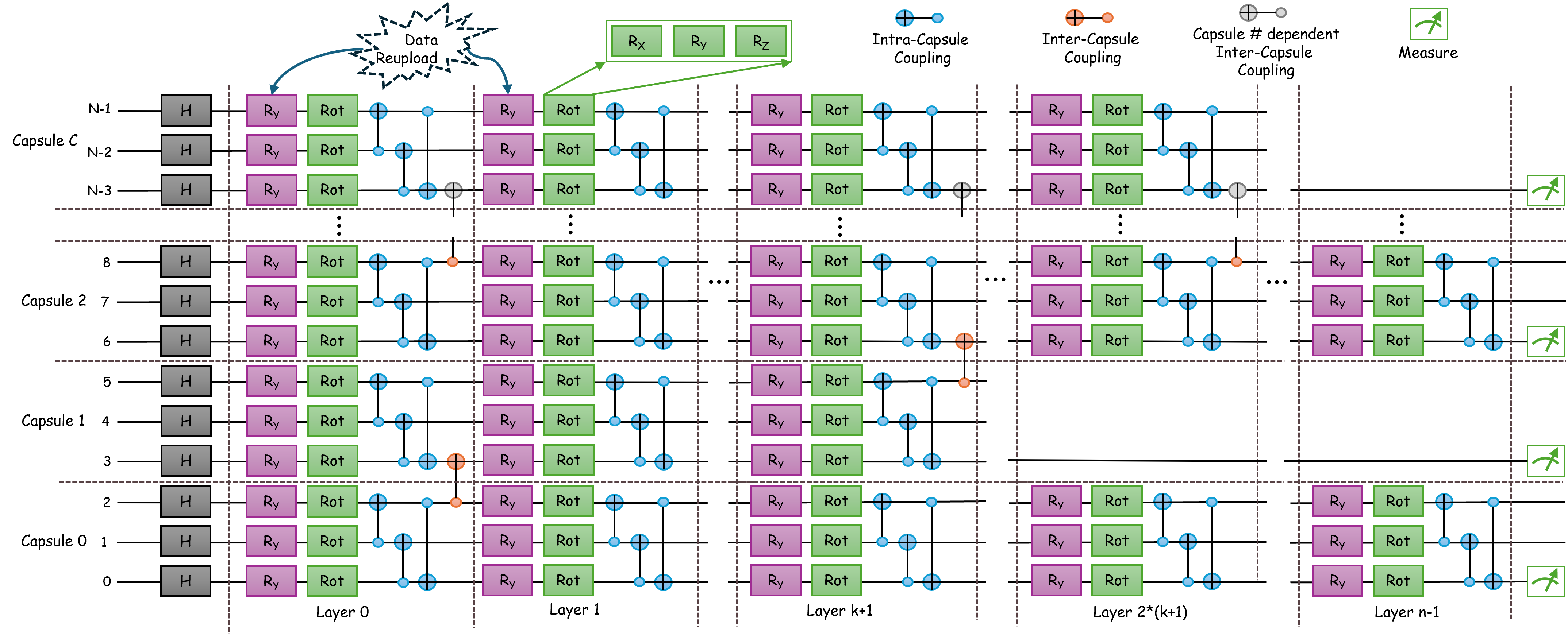}
    \caption{Overview of the Q-Capsule architecture. $N$ denotes the total number of qubits, $c$ the capsule index, $c_{\mathrm{size}}$ the number of qubits per capsule, and $N_c = N/c_{\mathrm{size}}$ the total number of capsules. $k$ denotes the inter-capsule coupling interval. $H$ is the Hadamard gate, $R_y$ performs trainable data re-uploading, and \textit{Rot} denotes a general trainable single-qubit rotation composed of $R_x$, $R_y$, and $R_z$. Blue control--target symbols denote intra-capsule CNOTs, orange symbols denote sparse inter-capsule CNOTs, gray symbols indicate capsule-dependent coupling determined by the alternating coupling pattern, and the green meter symbol denotes local capsule-wise measurement. Different capsule lengths illustrate independently adapted capsule depths.}
    \label{fig:sys_arch_qcapsule}
\end{figure}

\section{Methodology}
\label{sec:methodology}


\subsection{Capsule Design}
Local measurements are known to retain polynomially vanishing gradients in perceptron-based quantum networks~\cite{sharma2022trainability}. However, purely local measurements may fail to capture global correlations across the full quantum system, which can limit the information available for prediction. Inspired by this trade-off, we introduce qubit capsuling as a strategy to mitigate barren plateaus while preserving richer information within locally structured groups of qubits.

\textbf{Qubit Partitioning}
\label{sec:capsule}

For qubit partitioning, the $N$-qubit register is partitioned into $N_c=N/c_{\text{size}}$ disjoint
capsules of width $c_{\text{size}}$, capsule $c$ owning the contiguous qubits
$\mathcal{Q}_c=\{c\,c_{\text{size}},\dots,(c{+}1)c_{\text{size}}-1\}$. Each capsule maintains an independent depth $d_c$ and a private parameter
tensor $\bm{\theta}_c\in\mathbb{R}^{d_c\times c_{\text{size}}\times 4}$, laid
out contiguously in a flat parameter vector through a running offset table
$\{o_c\}$ with $o_{c+1}=o_c+4\,d_c\,c_{\text{size}}$. This layout enables heterogeneous capsule depths to grow independently, a single
capsule can be deepened (Section~\ref{sec:adaptive}) by extending its slice of
the vector without modifying the parameter allocation or offset structure of the remaining capsules.

\textbf{Local Readout and Per-Capsule Trainability}
\label{sec:readout}

Each capsule is measured using a local observable,
$O_c=Z_{c,c_{\text{size}}}$, producing one output value for each capsule,
$
\bm{o}(\bm{x};\bm{\theta})=
(\langle Z_{c,c_{\text{size}}}\rangle)_{c=0}^{N_c-1}
\in[-1,1]^{N_c}.
$
This local readout is an important design choice of our architecture. When the capsules are uncoupled ($k\to\infty$), or the circuit depth satisfies $L<k$, both the observable $O_c$ and the circuit operations that affect it are restricted to a capsule of fixed size $c_{\text{size}}$. The expectation $\langle O_c\rangle$ then depends only on the $c_{\text{size}}$ qubits of the capsule and their parameters. Even when the capsule sub-circuit forms a 2-design, the gradient variance of a single-qubit observable satisfies~\cite{mcclean2018barren}
\begin{equation}
\operatorname{Var}\!\left[\partial_{\theta}\langle O_c\rangle\right]
\;\in\;\Omega\!\left(2^{-c_{\text{size}}}\right)=\Omega(1)
\quad\text{in } N,
\label{eq:capsule_bound}
\end{equation}
and therefore does not decrease as the total number of qubits increases. With coupling every $k$ layers, the backward light cone of $O_c$ extends by one capsule on each side per coupling round. Its support then grows to $c_{\text{size}}\left(1+2\lfloor L/k\rfloor\right)$ qubits, and the bound becomes $\Omega\!\left(2^{-c_{\text{size}}(1+2\lfloor L/k\rfloor)}\right)$, which is still independent of $N$ for fixed $L$ and $k$.
However, a single global measurement can reintroduce barren plateaus even when the circuit is partitioned~\cite{cerezo2021cost}. Therefore, we use local capsule-wise measurements for classification and a global-$Z$ measurement only to analyze gradient concentration.

\subsection{Adaptive growth}
A quantum circuit with uniform depth may lead to inefficient allocation of computational resources, as feature blocks with higher complexity can become under-parameterized, while simpler blocks may receive more parameters than necessary. Q-Capsule addresses this limitation through an adaptive growth mechanism, allowing each capsule to expand independently according to its representational requirements. Specifically, capsule growth is guided by evidence of representational saturation measured using the Quantum Fisher Information Matrix (QFIM).

\textbf{QFIM-Based Expressibility Score}
\label{sec:qfim}
The adaptive controller requires a per-capsule criterion to determine whether a capsule has reached representational saturation at its current depth. We derive this criterion from the quantum Fisher information matrix (QFIM)~\cite{liu2019quantum}. For capsule $c$ with depth $d_c$, we evaluate the QFIM using an isolated $c_{\text{size}}$-qubit replica of the capsule sub-circuit under the block-diagonal metric-tensor approximation of Stokes~et~al.~\cite{stokes2020quantum}
\begin{equation}
\mathcal{F}_{ij}=4\,\mathrm{Re}\!\left[
\braket{\partial_i\psi_c|\partial_j\psi_c}
-\braket{\partial_i\psi_c|\psi_c}\!\braket{\psi_c|\partial_j\psi_c}\right],
\label{eq:qfim}
\end{equation}
Here the computation is performed on a $2^{c_{\text{size}}}$-dimensional state space and is therefore independent of total qubit $N$.

From the nonnegative, numerically clipped eigenvalue spectrum ${\lambda_i}$ of $F_c$, we define the expressibility score as
\begin{equation}
S_c=\frac{d_{\mathrm{eff}}(F_c)}
{\log_{10}\left(\kappa(F_c)+1\right)},
\qquad
d_{\mathrm{eff}}(F)=
\frac{\left(\sum_i\lambda_i\right)^2}
{\sum_i\lambda_i^2},
\qquad
\kappa(F)=
\frac{\lambda_{\max}}{\lambda_{\min}},
\label{eq:sscore}
\end{equation}
Motivated by the Fisher-geometric capacity measures introduced by Haug~et~al.~\cite{haug2021capacity}. Within this formulation, $d_{\mathrm{eff}}$ measures the number of independent and appreciably curved parameter-space directions available to the capsule, whereas $\kappa$ penalizes highly anisotropic spectra dominated by only a few stiff directions. 

\textbf{Saturation-Gated Adaptive Depth}
\label{sec:adaptive}

 Every $w_{\text{chk}}$ optimization steps, the controller appends the current $S_c$ to a per-capsule history and computes the windowed rate of change
 
 \begin{equation}
|\Delta S_c|=\frac{\bigl|S^{(t)}_c-S^{(t-w)}_c\bigr|}{w},
\label{eq:dS}
\end{equation}
over a window of length $w$. An increasing $S_c$ indicates that the capsule is still learning useful optimization directions, while a nearly constant $S_c$ suggests that its representational capacity has saturated at the current depth.
\begin{equation}
|\Delta S_c|<\tau \quad\text{and}\quad d_c<d_{\max},
\label{eq:growth_gate}
\end{equation}
 Here, $\tau$ denotes the saturation threshold that determines when the change in $S_c$ is sufficiently small to trigger circuit growth, while $d_{\max}$ specifies the maximum allowable depth of each capsule and prevents uncontrolled circuit expansion. Whenever an additional layer is appended only to capsule $c$, using the near-identity initialization strategy. Since the parameter vector changes dimensionality, the quantum node is reconstructed using the updated depth profile and the optimizer's moment estimates are reinitialized.

\subsection{Sparse Inter-Capsule Coupling and the Entanglement Budget}
\label{sec:sparse}

Since isolated capsules capture only local information, Q-Capsule introduces sparse coupling between neighboring capsules to model inter-block correlations. A CNOT gate connects the last qubit of capsule $c$ to the first qubit of capsule $c+1$ every $k$ layers using an alternating even--odd brickwork pattern, where $k$ denotes the \emph{coupling interval}. Thus, $k=1$ corresponds to coupling at every layer, whereas $k\rightarrow\infty$ yields fully isolated capsules.

We interpret the coupling interval as an entanglement regulator. Excessive entanglement between a local subsystem and the rest of the circuit can lead to concentration of the reduced state and vanishing local gradients~\cite{marrero2021entanglement}. Since each boundary CNOT can increase the entanglement across a capsule boundary by at most one ebit, the accumulated inter-capsule entanglement over depth $L$ is bounded as
\begin{equation}
S_{\text{inter}}(L)
\leq
\mathcal{O}\!\left(\left\lfloor \frac{L}{k}\right\rfloor\right).
\label{eq:ent_budget}
\end{equation}

More frequent coupling allows information to propagate across larger portions of the $N$-qubit register, increasing the risk of gradient concentration. We use $N/k$ as a simple measure of coupling density and consider the scaling relation
\begin{equation}
\operatorname{Var}\!\left[\partial_\theta\mathcal{L}\right]
\propto
\exp\!\left(-c^{*}\frac{N}{k}\right),
\qquad c^{*}>0.
\label{eq:scaling_law}
\end{equation}
This motivates choosing the coupling interval as $(k=\mathcal{O}(N))$, allowing the circuit to remain weakly coupled as the system size increases.

\subsection{Layer Structure and Trainable Re-Uploading}
\label{sec:layer}

The circuit begins with a global Hadamard layer that prepares the uniform superposition $|+\rangle^{\otimes N}$. In each layer $\ell<d_c$ of capsule $c$, there is a data-encoding layer followed by a trainable layer. So every qubit $j\in\mathcal{Q}_c$ undergo two rotations,
\begin{equation}
R_y\!\left(x_j\,\sigma^{(c)}_{\ell,j}\right)
\;\longrightarrow\;
\mathrm{Rot}\!\left(\alpha^{(c)}_{\ell,j},\beta^{(c)}_{\ell,j},
\gamma^{(c)}_{\ell,j}\right),
\label{eq:layer_gates}
\end{equation}
where $x_j$ is the input feature assigned to qubit $j$, and $\sigma^{(c)}_{\ell,j}$ is a trainable scaling parameter that controls how strongly the input feature is encoded at each layer. After these rotations, the qubits within each capsule are entangled using a nearest-neighbour CNOT ladder, with a ring connection when $c_{\text{size}}>2$. Importantly, no entangling gates connect qubits from different capsules in the base architecture.

The trainable encoding scale has an important role beyond simply adding parameters. Repeatedly encoding the input between trainable $SU(2)$ rotations allows each capsule to represent increasingly complex functions of its input features, with the accessible frequency spectrum growing as the number of data re-uploading layers increases. By learning $\sigma^{(c)}_{\ell,j}$, the circuit can adapt the strength and frequency of the encoded features to the data instead of using a fixed scale.


\section{Results}
\label{sec:results}

\subsection{Experimental setup}
\label{sec:config}
The overall experiments are conducted on two setups using the MNIST handwritten-digit dataset~\cite{lecun1998gradient}. For the binary classification task, two digit classes are selected, projected onto the first $N$ principal components, and min-max normalized to the interval $[0,\pi]$, such that each principal component is encoded into the rotation of a single qubit. For multiclass classification, we use the same experimental setup, except that the task involves four classes corresponding to digits 0--3. The resulting data are then divided into stratified training and test sets. In the classification pipeline, \textsc{Q-Capsule} acts as a quantum feature extractor in which the $N$ qubits are partitioned into $N_c$ capsules of size $c_{\text{size}}$. Each capsule processes its local subset of qubits through parameterized quantum layers while periodically exchanging information with neighboring capsules through sparse inter-capsule couplings. The features extracted from all capsules are then concatenated into a single feature vector and passed to a single classical output layer with a softmax activation to produce the final class prediction. Unless stated otherwise, the default experimental configuration uses $N=12$ qubits, $c_{\text{size}}=3$ (corresponding to $N_c=4$ capsules), coupling interval $k=3$, initial depth $d_{\text{init}}=2$, maximum depth $d_{\max}=8$, Adam learning rate $0.05$, growth window $w=8$, saturation threshold $\tau=0.02$, and a growth evaluation every six optimization steps over a training budget of $120$ iterations.

\subsection{Baseline}
We evaluate Q-Capsule against five baseline ans\"atze that together span the
principal strategies proposed for barren-plateau mitigation: a layerwise-grown
circuit (\textsc{Layer})~\cite{skolik2021layerwise}, a matrix-product-state
inspired circuit (\textsc{qMPS})~\cite{cervero2023barren}, a residual quantum network (\textsc{ResQNet})~\cite{kashif2024resqnets}, and two hardware-efficient
globally-entangling circuits with a global readout, denoted \textsc{Global} and \textsc{Glob96}~\cite{kandala2017hardware}. The comparison focuses on four key diagnostics: gradient-variance scaling with register width, gradient distributions at a fixed width, cost-landscape geometry, and parameter efficiency. Throughout, we interpret the results in terms of the mechanisms of that section rather than asserting performance in isolation, so that the empirical behaviour can be attributed to specific architectural choices.

\begin{figure}[htb!]
  \centering
  \includegraphics[width=\textwidth]{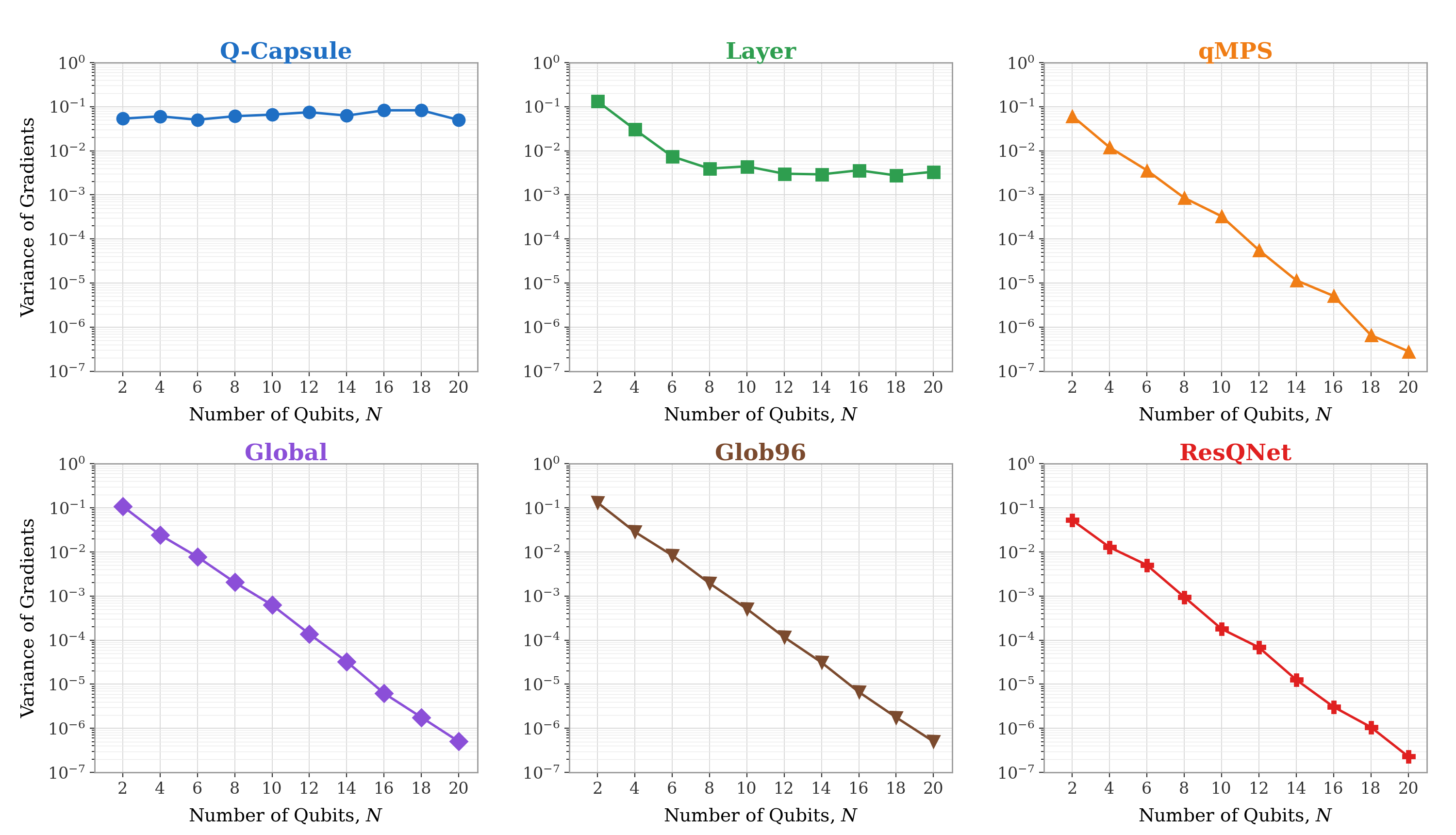}
  \caption{Gradient variance versus register width $N$ at fixed depth, for Q-Capsule and the five baselines (log scale).}
  \label{fig:exp1_scaling}
\end{figure}

\begin{figure}[htb!]
  \centering
  \includegraphics[width=\textwidth]{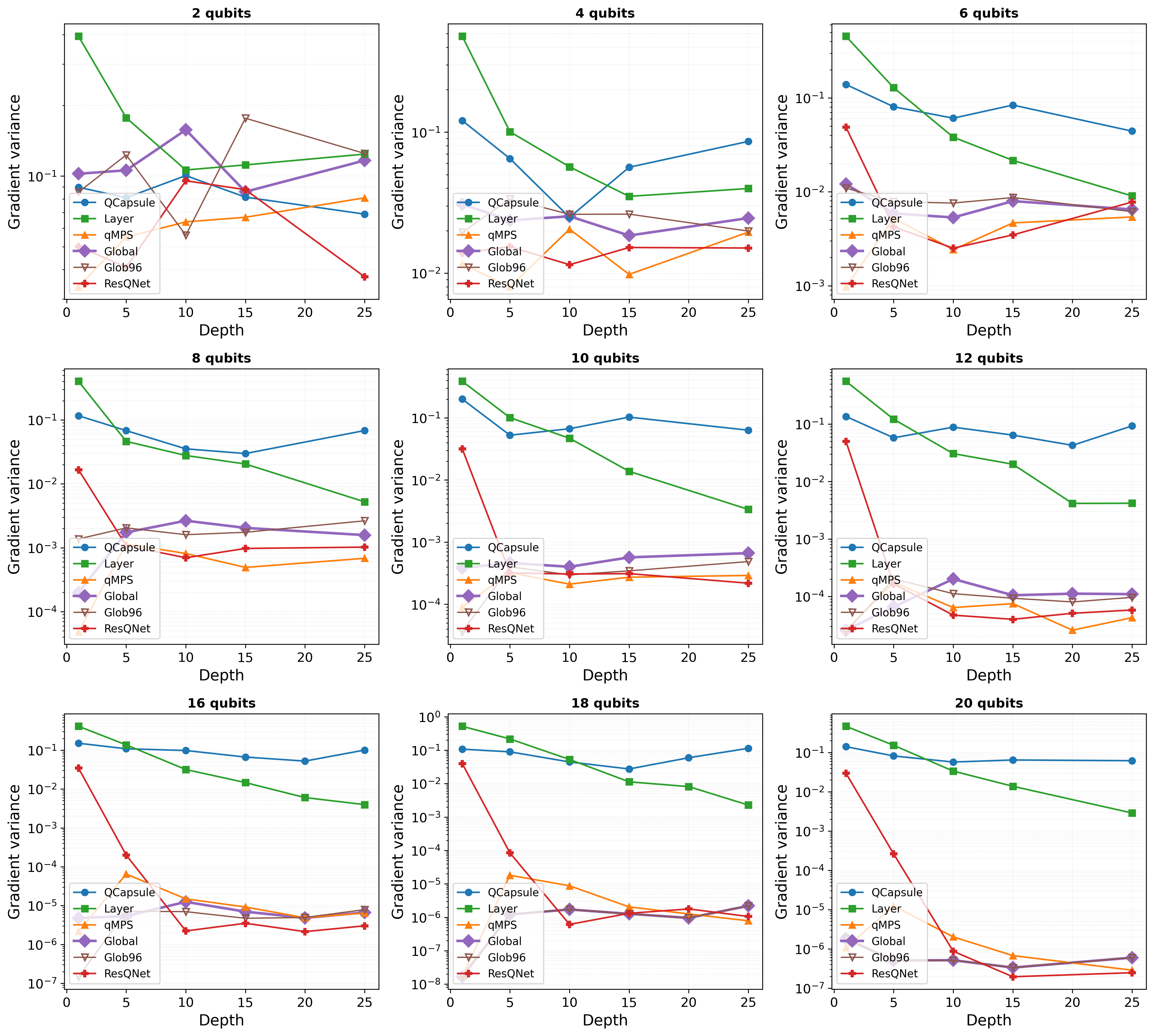}
  \caption{Gradient variance versus depth for each register width (log scale). Q-Capsule remains within an $\mathcal{O}(10^{-1})$ band across all depths.}
  \label{fig:exp3_depth}
\end{figure}

\subsection{Gradient Variance Scaling with Qubit Count and Circuit Depth}
\label{sec:res_scaling}

We evaluate the scalability of the gradient signal with respect to register width at a fixed circuit depth. Figure~\ref{fig:exp1_scaling} reports $\operatorname{Var}[\partial_{\theta}\mathcal{L}]$ for register widths $N\in{2,\ldots,20}$. The globally coupled architectures, e.g. \textsc{qMPS}, \textsc{Global}, \textsc{Glob96}, and \textsc{ResQNet}, exhibit a consistent exponential reduction in gradient variance as the number of qubits increases. For $N\geq4$, the fitted slopes of $\log\operatorname{Var}[\partial_{\theta}\mathcal{L}]$ are $-0.68$, $-0.69$, $-0.69$, and $-0.69$ per qubit, respectively. These values closely follow the characteristic $-\ln 2\approx-0.693$ scaling associated with exponential gradient suppression. In contrast, \textsc{Q-Capsule} maintains a gradient variance on the order of $10^{-1}$ across the full range of register widths, with a fitted slope of only $+0.01$ per qubit. The \textsc{Layer} baseline also avoids continued exponential decay and stabilizes near $10^{-3}$ at larger widths, although its gradient variance remains approximately one order of magnitude lower than that of \textsc{Q-Capsule}. Consequently, the separation between \textsc{Q-Capsule} and the globally coupled baselines grows with system size and reaches approximately five orders of magnitude at $N=20$. These results demonstrate that the localized capsule structure preserves strong gradient signals as the number of qubits increases.

We also examine the effect of circuit depth for $L\in{1,5,10,15,25}$ across different register widths. As shown in Figure~\ref{fig:exp3_depth}, \textsc{Q-Capsule} consistently maintains a gradient variance of approximately $10^{-1}$ as depth increases. The globally entangling baselines, which are already suppressed at larger register widths, exhibit an additional reduction in gradient variance at greater depths.  The depth stability of \textsc{Q-Capsule} indicates that increasing circuit depth does not lead to substantial gradient suppression within the evaluated range. This result suggests that sparse inter-capsule coupling constrains excessive global entanglement while allowing additional layers to increase the representational capacity within individual capsules. The \textsc{Global} and \textsc{Glob96} baselines exhibit nearly identical behavior at larger register widths, with their curves almost overlapping for $N=18$ and $N=20$. Thus, differences in their parameterization have little effect once the circuits enter a strongly suppressed gradient regime.

\begin{figure}[htb!]
  \centering
  \includegraphics[width=\textwidth]{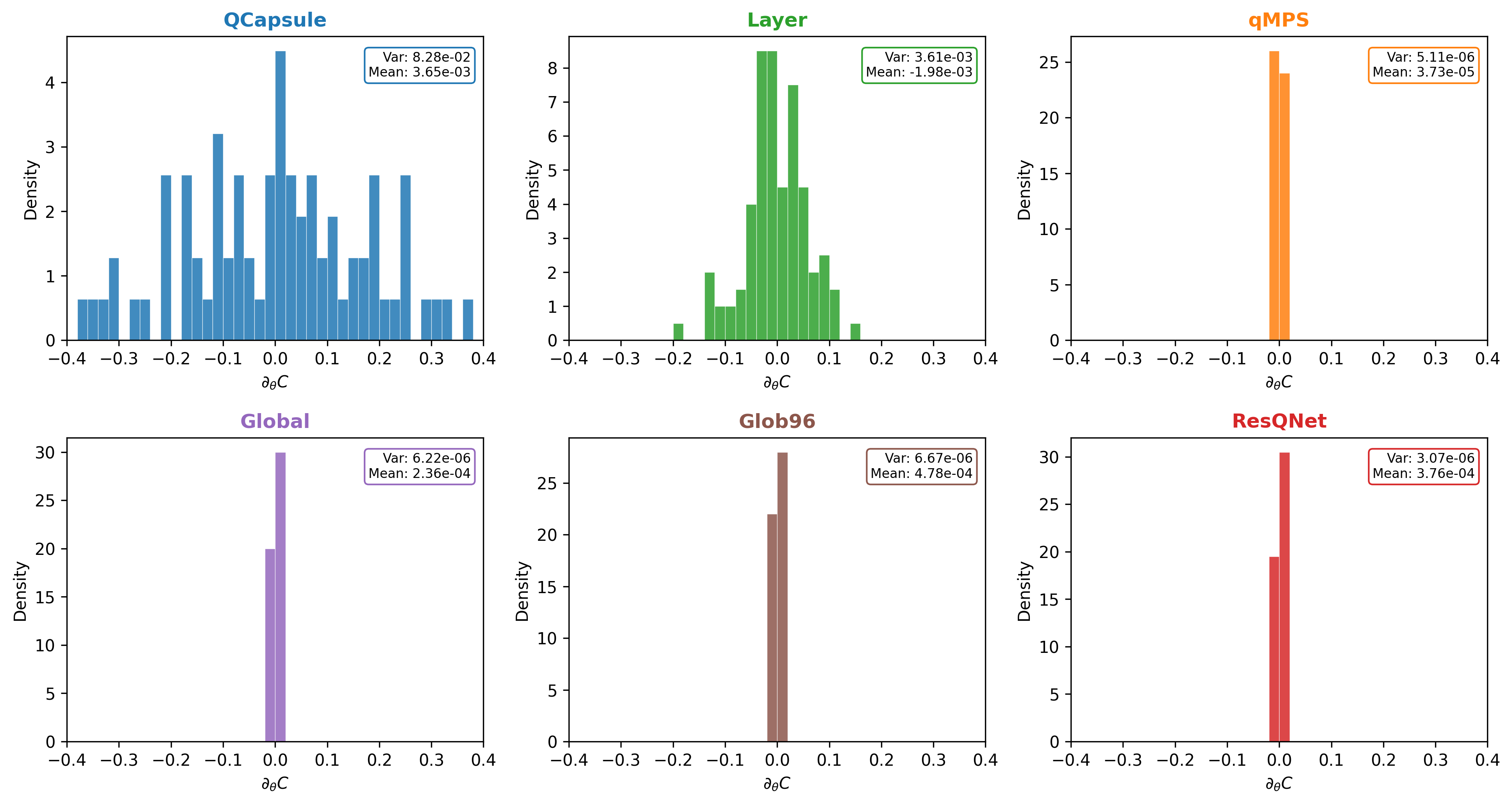}
  \caption{Empirical gradient distributions at $N=16$, max depth $25$, over random
  initializations, with per-model variance and mean annotated.}
  \label{fig:exp2_hist}
\end{figure}

\subsection{Gradient Distributions at Fixed Width}
\label{sec:res_hist}

Aggregation of variance can hide the shape of the underlying gradient distribution. Figure~\ref{fig:exp2_hist} shows the distribution of $\partial_\theta\mathcal{L}$ over an ensemble of random initializations at a representative large register ($N=16$, depth $25$), with the per-model empirical variance annotated. To account for finite-shot measurement variability, each gradient estimate is evaluated using $100$ shots, and the reported values represent the corresponding average estimates. The distributions make the width-scaling result concrete at the level of a single operating point. The Q-Capsule gradients are broadly distributed over the interval $[-0.5,0.5]$, with an empirical variance of $8.3\times10^{-2}$ and individual components reaching magnitudes above $0.7$, indicating that typical descent directions remain well separated from zero. The globally-read baselines, in contrast, are sharply concentrated around the origin: \textsc{Global}, \textsc{Glob96}, \textsc{qMPS}, and \textsc{ResQNet} all exhibit variances between $3\times10^{-6}$ and $7\times10^{-6}$, roughly four orders of magnitude below Q-Capsule, with most of their gradient mass confined to a narrow region near zero. The \textsc{Layer} baseline is again intermediate, with a variance of $3.6\times10^{-3}$. Because the number of measurement shots required to reliably resolve a descent direction scales inversely with the gradient variance, the four-order-of-magnitude separation between Q-Capsule and the globally-read baselines implies a comparable difference in measurement cost at this width. This provides a practical interpretation of the barren plateau: the baseline gradients become increasingly difficult to resolve, whereas Q-Capsule retains gradients of sufficient magnitude for effective optimization.

\begin{figure}[htb!]
  \centering
  \includegraphics[width=\textwidth]{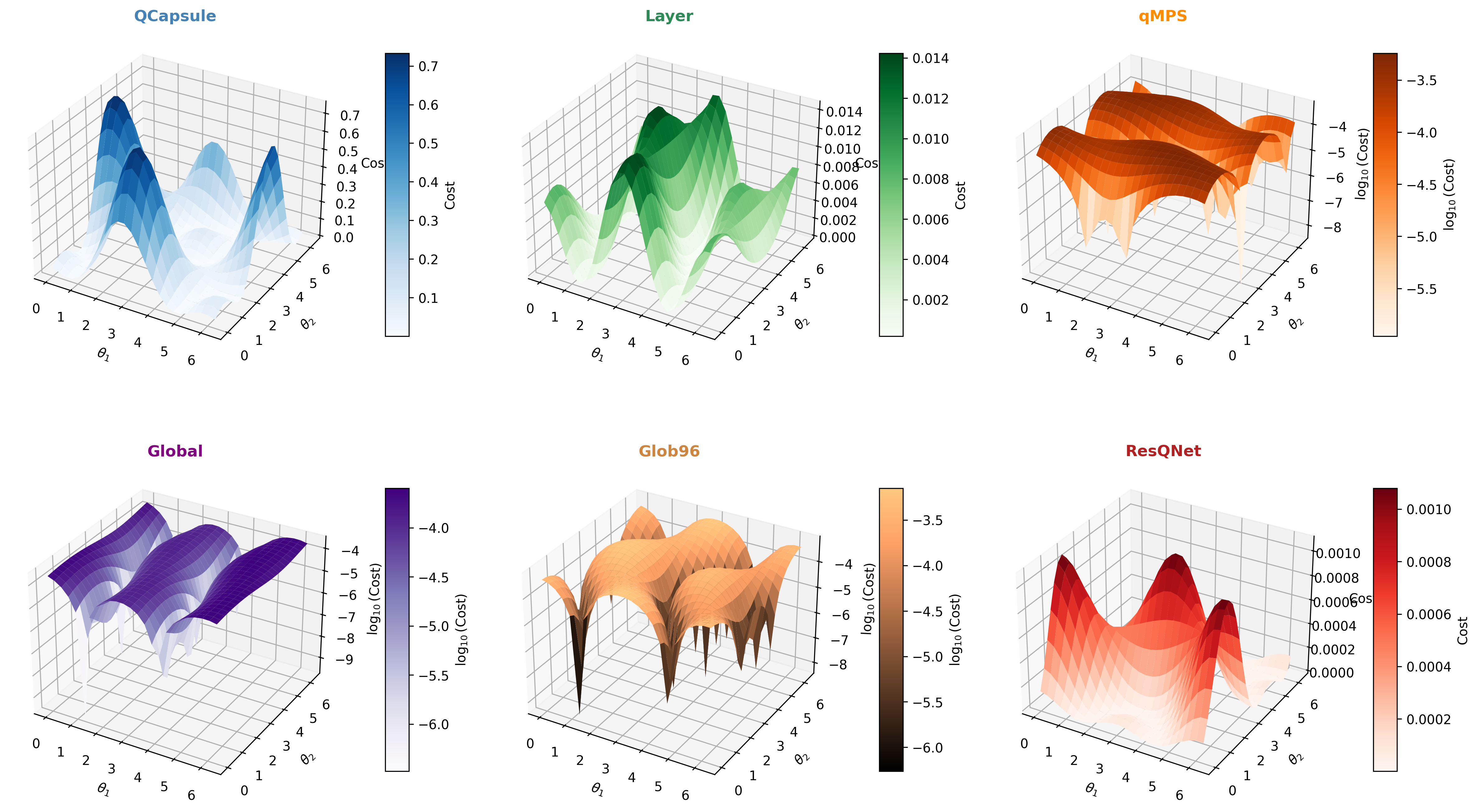}
  \caption{Two-parameter cost-landscape slices $(\theta_1,\theta_2)$ at fixed
  remaining parameters. Q-Capsule (linear colour scale) shows structured minima
  spanning a cost range $\approx0.75$.}
  \label{fig:exp4_landscape}
\end{figure}

\subsection{Cost-Landscape Geometry}
\label{sec:res_landscape}

To further characterize the optimization geometry of different barren Plateau Mitigation techniques, we evaluate the loss over a two-parameter slice $(\theta_1,\theta_2)\in[0,2\pi)^2$, while keeping all remaining parameters fixed across architectures. Figure~\ref{fig:exp4_landscape} shows that \textsc{Q-Capsule} produces a structured landscape with distinct variations and well-separated minima. Its loss spans a total range of approximately $0.75$, providing clear directions for gradient-based optimization. In contrast, the globally coupled baselines produce nearly flat landscapes. The total loss variation of \textsc{Global} is approximately $2.6\times10^{-4}$, while \textsc{qMPS}, \textsc{Glob96}, and \textsc{ResQNet} exhibit similarly limited variation. The approximately three-order-of-magnitude difference in landscape variation is consistent with the gradient-variance results. Together, these findings show that \textsc{Q-Capsule} preserves meaningful local cost variations, whereas the globally coupled architectures operate in regions that are effectively flat at the evaluated width and depth.

\begin{figure*}[htb!]
  \centering

  \begin{minipage}[htb!]{0.45\textwidth}
    \centering
    \includegraphics[width=\linewidth]{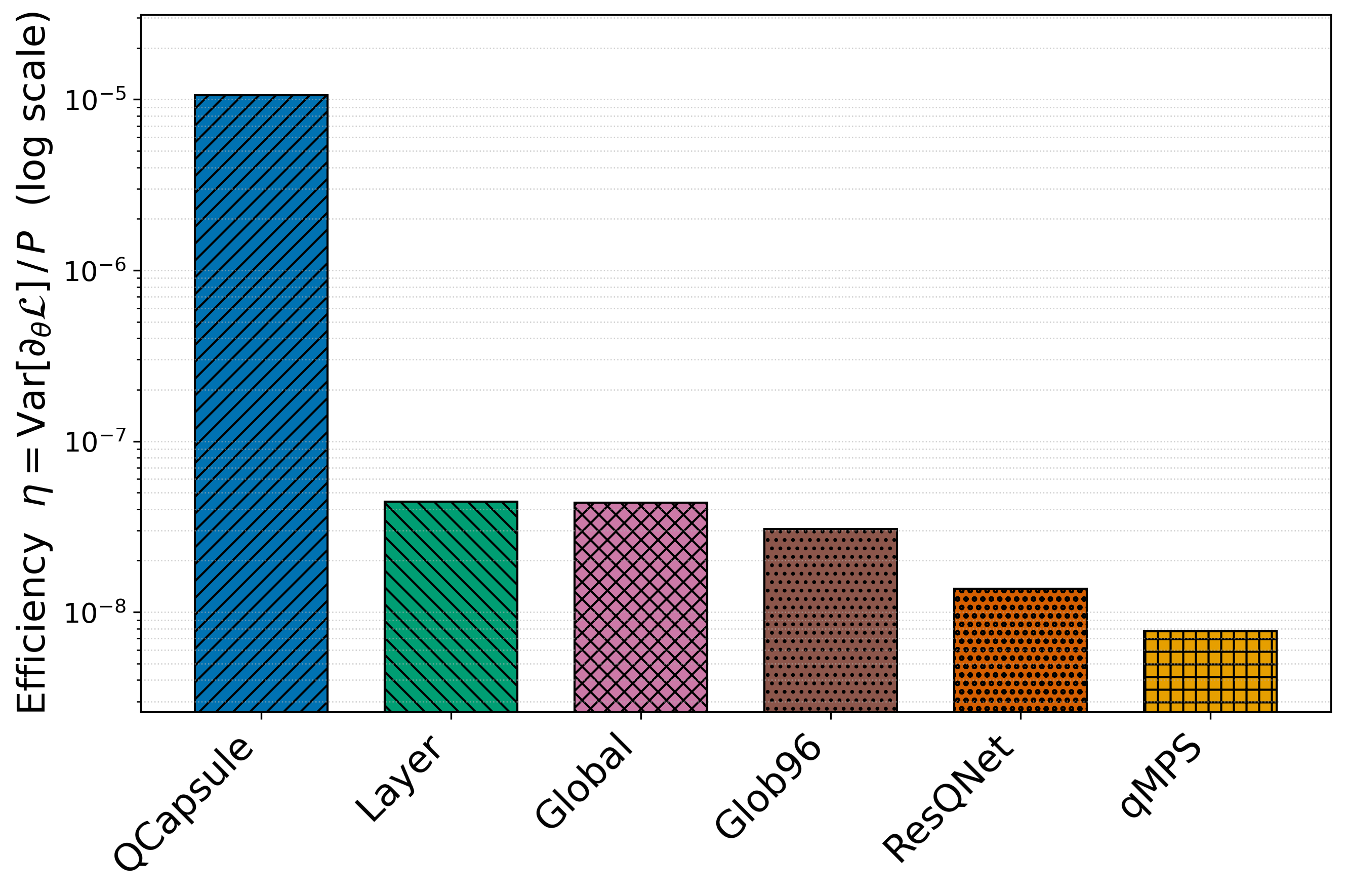}
    \captionof{figure}{Parameter-efficiency score per model (log scale).
    Q-Capsule exceeds the next-best baselines by approximately two orders of
    magnitude, consistent with the saturation-gated allocation of
    Equation~\eqref{eq:growth_gate}.}
    \label{fig:exp_efficiency}
  \end{minipage}
  \hfill
  \begin{minipage}[htb!]{0.5\textwidth}
  \centering
  \captionof{table}{Parameter-efficiency score
  $\eta=\operatorname{Var}[\partial_{\theta}\mathcal{L}]/P$ at $N=12$.
  The final column reports its advantage relative to each baseline.}
  \label{tab:res_efficiency}
  \setlength{\aboverulesep}{0pt}
  \setlength{\belowrulesep}{0pt}
  \setlength{\extrarowheight}{2pt}
  \resizebox{\linewidth}{!}{%
  \begin{tabular}{l
                  >{\columncolor{colgreen}}c
                  >{\columncolor{colcream}}c
                  >{\columncolor{colmauve}}c
                  >{\columncolor{colsteel}}c}
    \toprule
    & \cellcolor{hdrgreen}\textbf{Gradient}
    & \cellcolor{hdrcream}\textbf{Params}
    & \cellcolor{hdrmauve}
    & \cellcolor{hdrsteel}\textbf{Rel.\ to} \\
    \multirow{-2}{*}{\textbf{Model}}
    & \cellcolor{hdrgreen}\textbf{variance}
    & \cellcolor{hdrcream}\textbf{$P$}
    & \cellcolor{hdrmauve}\multirow{-2}{*}{$\boldsymbol{\eta}$}
    & \cellcolor{hdrsteel}\textbf{Q-Capsule} \\
    \midrule
    \rowcolor{rowhilite}\textbf{\textsc{Q-Capsule}} & $\mathbf{5.08\times10^{-2}}$ & $4800$
      & $\mathbf{1.06\times10^{-5}}$ & $1.0\times$ \\
    \textsc{Layer} & $1.07\times10^{-4}$ & $2400$
      & $4.44\times10^{-8}$ & $238\times$ \\
    \textsc{Global} & $1.05\times10^{-4}$ & $2400$
      & $4.36\times10^{-8}$ & $243\times$ \\
    \textsc{Glob96} & $1.47\times10^{-4}$ & $4800$
      & $3.07\times10^{-8}$ & $345\times$ \\
    \textsc{ResQNet} & $6.59\times10^{-5}$ & $4800$
      & $1.37\times10^{-8}$ & $771\times$ \\
    \textsc{qMPS} & $6.80\times10^{-5}$ & $8800$
      & $7.73\times10^{-9}$ & $1370\times$ \\
    \bottomrule
  \end{tabular}%
  }
\end{minipage}

\end{figure*}

\subsection{Parameter Efficiency}
\label{sec:res_efficiency}

A trainable architecture should preserve useful gradients without relying on an inflated parameter budget. We quantify this trade-off using the
parameter-efficiency score $\eta \;=\; \frac{\operatorname{Var}[\partial_{\theta}\mathcal{L}]}{P}$ where $P$ denotes the number of trainable parameters.  The score measures the amount of gradient signal retained per trainable parameter and is used only for relative comparison across architectures. As shown in Figure~\ref{fig:exp_efficiency} and
Table~\ref{tab:res_efficiency}, \textsc{Q-Capsule} achieves the highest efficiency, $\eta=1.06\times10^{-5}$, exceeding the next-best baselines by more than two orders of magnitude and the least efficient baseline by nearly three. The two hardware-efficient globally coupled circuits, \textsc{Global} and \textsc{Layer}, follow at $\eta=4.36\times10^{-8}$ and $4.44\times10^{-8}$
respectively---roughly $240\times$ below \textsc{Q-Capsule}---while
\textsc{Glob96} ($3.07\times10^{-8}$), \textsc{ResQNet} ($1.37\times10^{-8}$),
and \textsc{qMPS} ($7.73\times10^{-9}$) fall further behind, the last by a factor
of approximately $1.4\times10^{3}$.

This advantage is driven primarily by the substantially larger gradient variance maintained by \textsc{Q-Capsule} ($5.08\times10^{-2}$), against
$\mathcal{O}(10^{-4})$--$\mathcal{O}(10^{-5})$ for the baselines. Rather than by a reduction in model capacity: \textsc{Q-Capsule} in fact carries $4800$ trainable parameters at $N=12$ so its ranking cannot be attributed to a smaller parameter budget. The result therefore confirms that the improved trainability of \textsc{Q-Capsule}
is not obtained through parameter inflation. It is also consistent with the
adaptive growth mechanism, which introduces additional capacity only after the
existing capsule structure approaches saturation.

\subsection{Ablation Study}
\label{sec:ablation}
We conduct a series of ablation experiments to isolate the contributions of capsule partitioning, capsule size, sparse inter-capsule coupling, data re-uploading, readout locality, and structured initialization. We report both gradient variance and classification accuracy, as these quantities characterize complementary aspects of model trainability.

\subsubsection{Capsule Partitioning and Capsule Size}
\label{sec:abl_capsule}
\begin{figure*}
\centering

\begin{minipage}[htb!]{0.47\textwidth}
  \vspace{0pt}
  \centering
  \includegraphics[width=\linewidth]{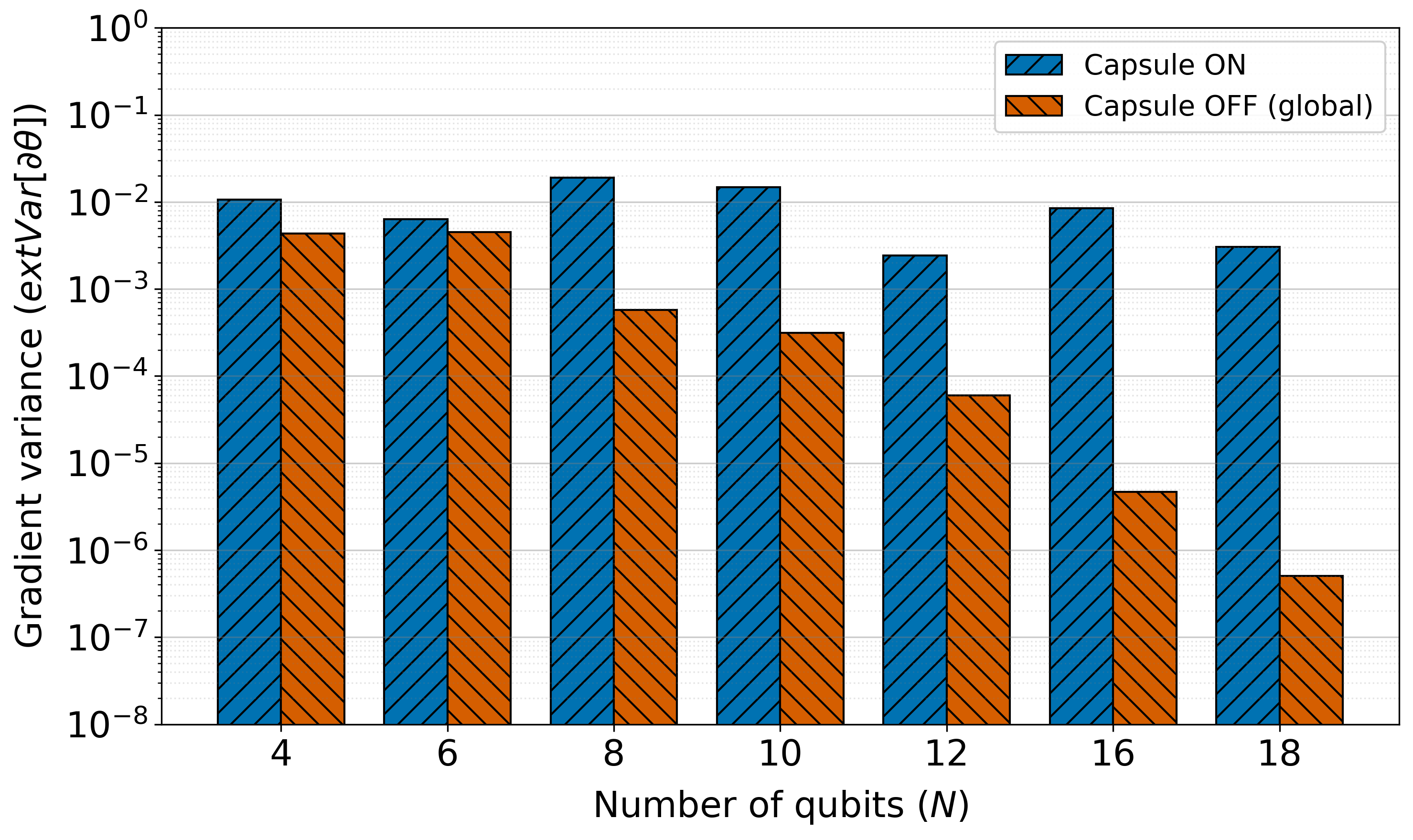}

  \captionsetup{type=figure,skip=3pt}
  \captionof{figure}{Gradient variance with and without capsule partitioning as the number of qubits increases.}
  \label{fig:abl_capsule}
\end{minipage}
\hfill
\begin{minipage}[htb!]{0.49\textwidth}
  \vspace{0pt}
  \centering
  \captionsetup{type=table,skip=4pt}
  \captionof{table}{Influence of capsule size on the number of capsules, gradient variance, and classification accuracy at $N=12$. Bold marks the configuration used throughout.}
  \label{tab:abl_capsize}
  \setlength{\aboverulesep}{0pt}
  \setlength{\belowrulesep}{0pt}
  \setlength{\extrarowheight}{2pt}
  \resizebox{\linewidth}{!}{%
    \begin{tabular}{c
                    >{\columncolor{colcream}}c
                    >{\columncolor{colgreen}}c
                    >{\columncolor{colsteel}}c}
      \toprule
      \textbf{$c_{\text{size}}$}
      & \cellcolor{hdrcream}\textbf{Capsules} $\boldsymbol{N_c}$
      & \cellcolor{hdrgreen}\textbf{Gradient variance}
      & \cellcolor{hdrsteel}\textbf{Accuracy} \\
      \midrule
      $2$ & $6$ & $1.45\times10^{-2}$ & $0.954$ \\
      $\mathbf{3}$ & $4$ & $2.42\times10^{-3}$ & $\mathbf{0.972}$ \\
      $4$ & $3$ & $1.51\times10^{-3}$ & $0.963$ \\
      $6$ & $2$ & $2.37\times10^{-4}$ & $0.954$ \\
      \bottomrule
    \end{tabular}%
  }
\end{minipage}

\end{figure*}

We first evaluate the role of capsule partitioning by replacing the
capsule-structured architecture with a single globally entangling circuit and a global readout. As shown in Figure~\ref{fig:abl_capsule}, the global variant exhibits pronounced gradient suppression as the register width increases. Its gradient variance decreases from $4.3\times10^{-3}$ at $N=4$ to $5.1\times10^{-7}$ at $N=18$. In contrast, the capsule-partitioned model maintains gradient variance between $\mathcal{O}(10^{-3})$ and $\mathcal{O}(10^{-2})$ over the same range, without a systematic decay with $N$. Consequently, the separation between the two variants grows with register width and exceeds three orders of magnitude at $N=18$. These results identify capsule partitioning as a primary factor in preventing width-induced gradient suppression.

We next examine the effect of capsule size by varying $c_{\text{size}}\in\{2,3,4,6\}$ at $N=12$. As reported in Table~\ref{tab:abl_capsize}, the gradient variance decreases monotonically from $1.45\times10^{-2}$ at $c_{\text{size}}=2$ to $2.37\times10^{-4}$ at $c_{\text{size}}=6$, corresponding to a reduction of nearly two orders of magnitude. This trend is consistent with the $\Omega(2^{-c_{\text{size}}})$ lower bound: as each capsule spans a larger Hilbert space, its local gradients become increasingly susceptible to suppression. Accuracy exhibits a non-monotonic dependence on capsule size. The highest accuracy, $0.972$, is obtained at $c_{\text{size}}=3$. The smaller $c_{\text{size}}=2$ configuration retains stronger gradients but achieves lower accuracy, suggesting insufficient within-capsule expressivity. Conversely, larger capsules provide greater local capacity but exhibit substantially weaker gradients. The default choice $c_{\text{size}}=3$ therefore provides the most favorable balance between predictive performance and gradient preservation.



\subsubsection{Information Shared by Sparse Coupling}
\label{sec:perf_coupling_info}

Sparse inter-capsule coupling preserves gradient magnitude, but it may also
reduce the information shared between neighbouring capsules. To examine this
trade-off we measure the quantum mutual information between two adjacent
capsules, $I(A{:}B)=S(A)+S(B)-S(AB)$, where $S(\cdot)$ denotes the von Neumann
entropy; a value of zero indicates that the two capsules are statistically
independent. Figure~\ref{fig:perf_coupling} reports $I(A{:}B)$ together with the
number of boundary CNOT gates that each coupling interval $k$ introduces.

In the isolated configuration the mutual information is exactly $0.00$ bits.
This confirms that the capsules are strictly independent when the boundary gates
are removed, and therefore that any information observed at the remaining
settings is attributable to those gates alone rather than to the encoding or the
intra-capsule structure. Once coupling is enabled, the two capsules share between
$1.64$ and $3.06$ bits---a substantial portion of the $6$-bit maximum available
to a pair of three-qubit capsules.

The central observation is that this shared information does not scale with the
number of coupling gates. The densest setting, $k=1$, spends nine boundary CNOTs
and reaches $2.29$ bits, whereas $k=2$ spends only five and attains the highest
value recorded, $3.06$ bits. At the sparsest setting tested, $k=8$, two CNOTs
still sustain $1.92$ bits, i.e.\ $84\%$ of the information obtained by the
densest configuration at $22\%$ of its coupling cost. Information sharing is
therefore not purchased gate by gate: beyond the small number of boundary gates
required to correlate the two capsules at all, additional coupling contributes
little.

That the densest setting is not the most informative is consistent with the
monogamy of entanglement. At $k=1$ every capsule boundary is coupled at every
layer, so correlations established at one boundary are redistributed across the
full four-capsule register; the entanglement available to any particular pair is
correspondingly reduced. Intermediate coupling confines the correlations to
neighbouring capsules, which is precisely where the mutual information is
measured. The quantity relevant to the architecture is thus how the entanglement
budget is allocated, not how much of it is generated.

Taken together with circular intra-capsule couplin, these measurements delineate
the operating regime of the capsule structure. Reducing coupling strengthens the
local gradient signal, and the present results show that it does so without
severing communication between capsules: only the fully decoupled configuration
produces genuinely independent capsules, while every sparse setting retains most
of the shared information at a fraction of the two-qubit gate cost. Sparse
coupling therefore preserves gradient magnitude and inter-capsule communication
simultaneously, rather than trading one against the other.

\begin{figure*}
\centering

\begin{minipage}[htb!]{0.47\textwidth}
  \vspace{0pt}
  \centering
  \includegraphics[width=\linewidth]{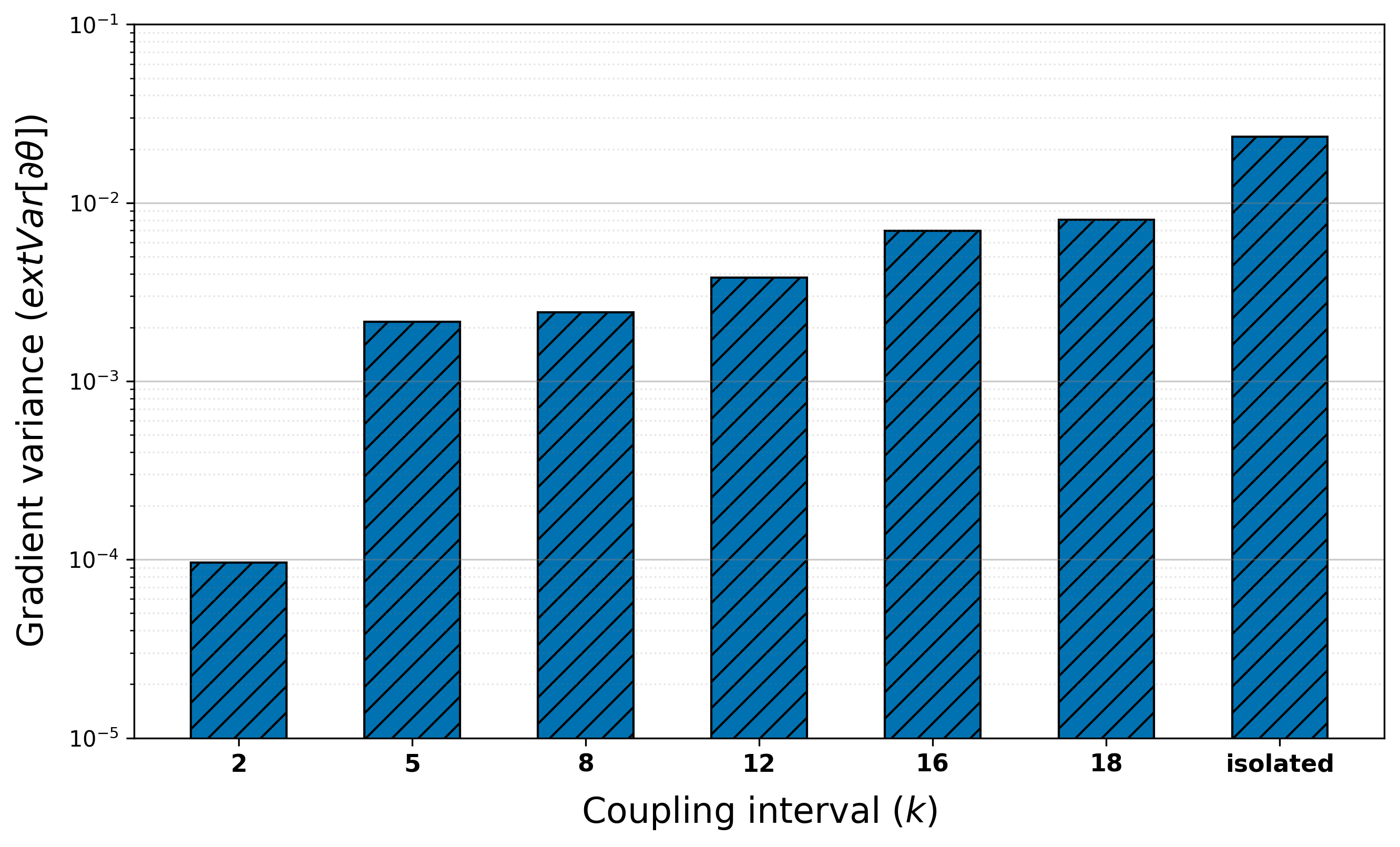}

  \captionsetup{type=figure,skip=3pt}
  \captionof{figure}{Effect of inter-capsule coupling interval k on gradient variance, improved gradient preservation with sparser coupling.}
  \label{fig:abl_coupling}
\end{minipage}
\hfill
\begin{minipage}[htb!]{0.49\textwidth}
  \vspace{0pt}
  \centering
  \captionsetup{type=table,skip=4pt}
  \captionof{table}{Effect of key architectural components on gradient variance and classification accuracy. Bold settings denote the full model.}
  \label{tab:abl_components}
  \setlength{\aboverulesep}{0pt}
  \setlength{\belowrulesep}{0pt}
  \setlength{\extrarowheight}{2pt}
  \resizebox{\linewidth}{!}{%
    \begin{tabular}{l
                    >{\columncolor{colcream}}l
                    >{\columncolor{colgreen}}c
                    >{\columncolor{colsteel}}c}
    \toprule
    \textbf{Component}
      & \cellcolor{hdrcream}\textbf{Setting}
      & \cellcolor{hdrgreen}\textbf{Gradient variance}
      & \cellcolor{hdrsteel}\textbf{Accuracy} \\
    \midrule
      & \textbf{Enabled (full)} & $2.42\times10^{-3}$ & $\mathbf{0.972}$ \\
    \multirow{-2}{*}{Data re-uploading}
      & Removed                 & $1.30\times10^{-2}$ & $0.509$ \\
    \midrule
      & \textbf{Local $Z_0$ (full)} & $2.42\times10^{-3}$ & $\mathbf{0.972}$ \\
    \multirow{-2}{*}{Readout observable}
      & Global $Z_0Z_{n-1}$         & $9.77\times10^{-5}$ & $0.963$ \\
    \midrule
      & \textbf{Structured (full)}  & $2.42\times10^{-3}$ & $\mathbf{0.972}$ \\
    \multirow{-2}{*}{Initialization}
      & Uniform $[0,2\pi)$          & $1.99\times10^{-3}$ & $0.639$ \\
    \bottomrule
  \end{tabular}
  }
\end{minipage}

\end{figure*}

\subsubsection{Component Knock-Out: Re-Uploading, Readout Locality, Initialization}
\label{sec:abl_components}

To understand the importance of  trainable data re-uploading, local readout, and structured initialization we evaluated the Q-Capsule with and without three components. Table~\ref{tab:abl_components} reports the resulting gradient variance and classification accuracy. Removing data re-uploading causes accuracy to collapse from $0.972$ to $0.509$, which is approximately chance performance on the balanced binary task. Notably, the gradient variance increases to $1.30\times10^{-2}$. Thus, the failure cannot be attributed to vanishing gradients. Instead, the result indicates that re-uploading is required to maintain an effective pathway through which the classical input influences the learned quantum representation. Removing structured initialization similarly reduces accuracy, from $0.972$ to $0.639$, while leaving the gradient variance nearly unchanged at $1.99\times10^{-3}$. This result suggests that gradient magnitude alone does not characterize whether the available descent directions are aligned with useful data-dependent features. Structured initialization therefore contributes not only by preserving gradient scale, but also by placing the circuit in a parameter regime from which effective representations can be learned.

A different pattern emerges when the local observable is replaced with the
global two-qubit observable $Z_0Z_{N-1}$. In this case, the gradient variance decreases from $2.42\times10^{-3}$ to $9.77\times10^{-5}$, whereas accuracy decreases only slightly from $0.972$ to $0.963$. The reduction in variance is consistent with the cost-locality analysis \cite{cerezo2021cost}, according to which global observables are more
susceptible to barren plateaus than local observables. The limited accuracy
reduction is likely attributable to the moderate register width used in this
experiment. The trainability disadvantage of the global readout is expected to
become more consequential as register size increases.

\begin{figure*}[htb!]
  \centering

  \begin{minipage}[t]{0.59\textwidth}
    \centering
    \includegraphics[width=\linewidth]{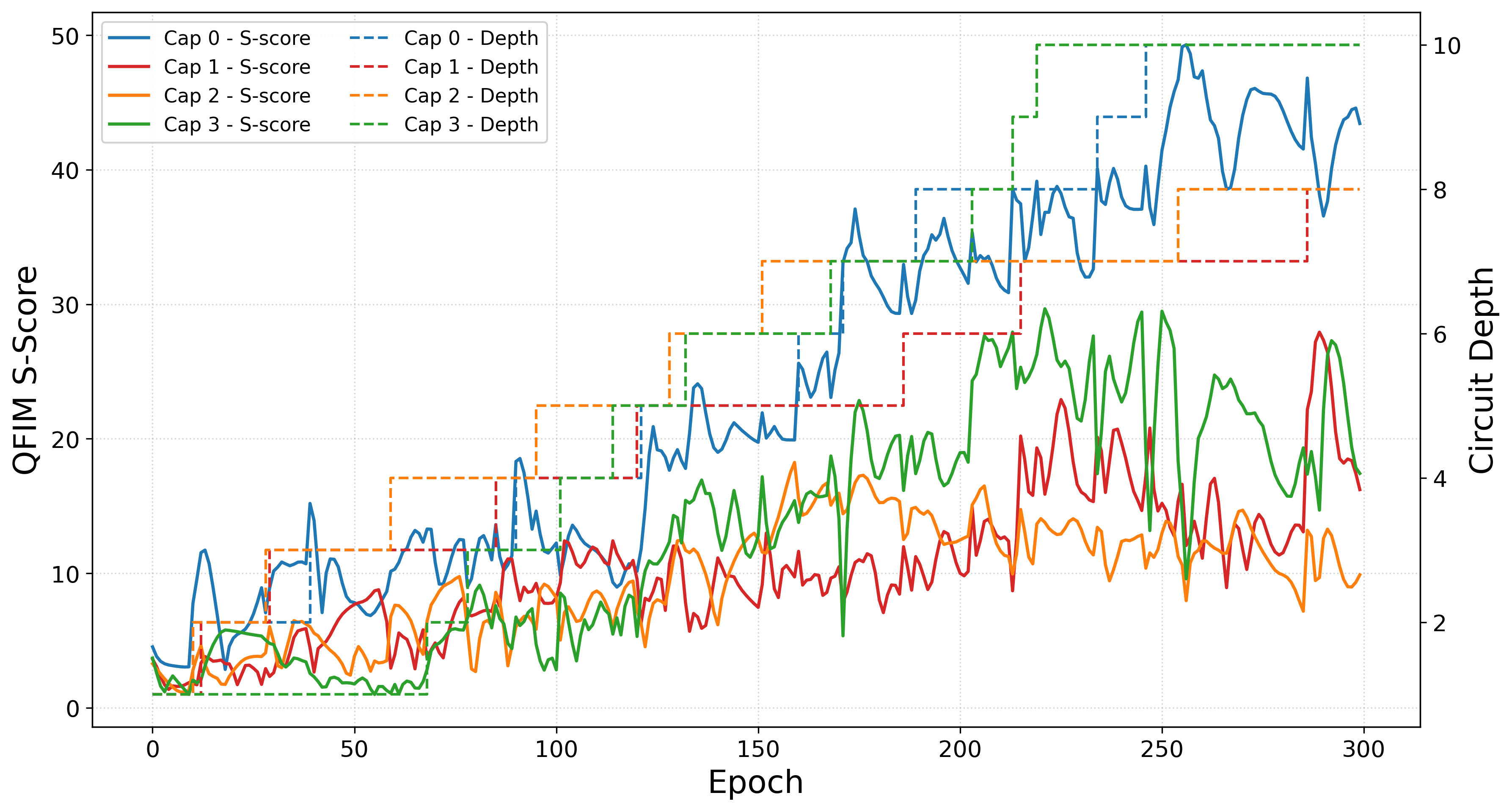}
    \captionof{figure}{QFIM S-score (solid, left axis) and capsule depth
    (dashed step, right axis) versus training epoch for the four capsules of
    the default configuration. Each depth increment driven by the growth gate
    of Equation~\eqref{eq:growth_gate} is followed by a sustained rise in
    S-score.}
    \label{fig:perf_sscore}
  \end{minipage}
  \hfill

\end{figure*}
\subsubsection{Adaptive Growth of Representational Capacity}
\label{sec:perf_adaptive_growth}

To determine whether this added depth provides useful capacity, we track the effective dimension of each capsule using the QFIM-based S-score till  $d_{\max}=10$. Figure~\ref{fig:perf_sscore} shows that the S-score increases with capsule depth. Each depth increase is followed by a sustained rise in S-score, from approximately $3$ at $d=1$ to about $40$ for the deepest capsule. Across all capsules, the mean S-score increases from $2.7$ at $d=1$ to $13.9$ at $d=6$ and approximately $31$ at $d=10$, with correlations between depth and S-score ranging from $0.85$ to $0.98$. These results indicate that adaptive growth increases effective representational capacity rather than adding inaccessible parameters.

\begin{figure*}[htb!]
  \centering

  \begin{minipage}[htb!]{0.48\textwidth}
    \centering
    \vspace{0pt}
    \includegraphics[width=\linewidth]{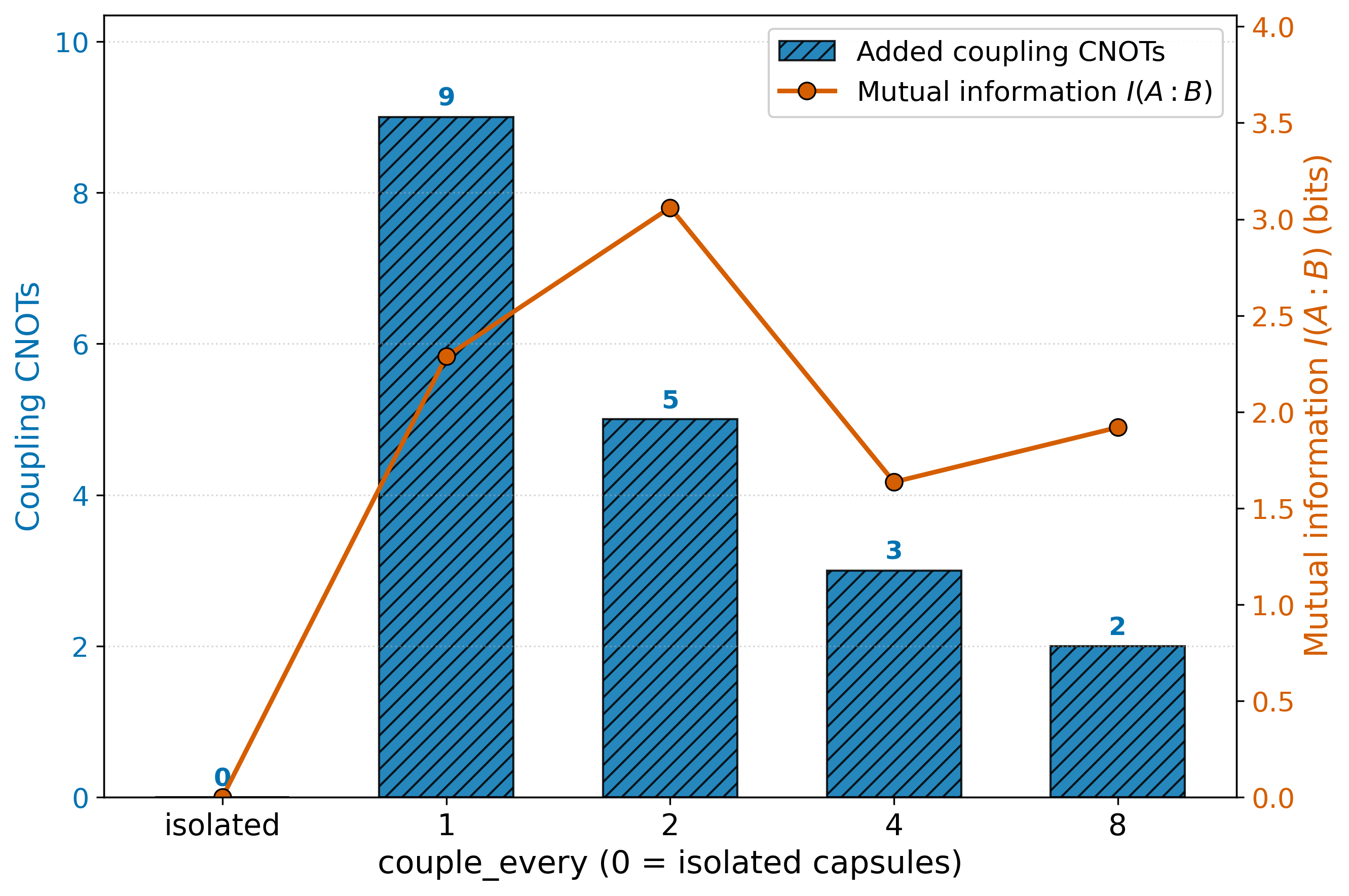}
    \captionof{figure}{Quantum mutual information $I(A{:}B)$ between two adjacent
    capsules at $N=12$ versus the coupling spacing, together with the added
    boundary-CNOT count (left axis). The isolated limit shares $0$ bits,
    while sparse coupling retains most of the shared information at a
    fraction of the coupling-gate cost.}
    \label{fig:perf_coupling}
  \end{minipage}
  \hfill
  \begin{minipage}[htb!]{0.48\textwidth}
    \centering
    \vspace{0pt}
    \includegraphics[width=\linewidth]{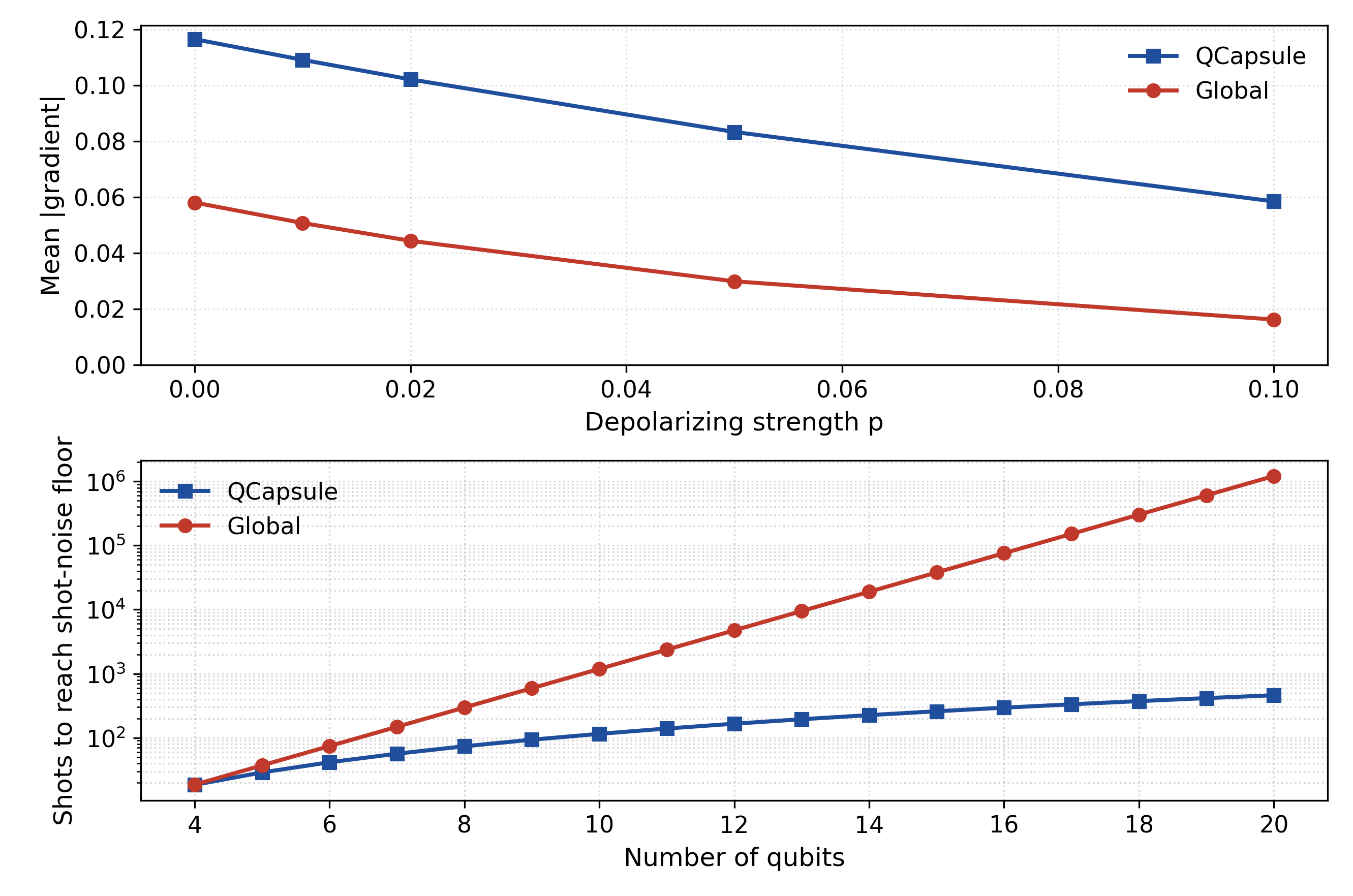}
    \captionof{figure}{Top: mean gradient magnitude under a per-layer
    depolarizing channel of strength $p$. The Q-Capsule signal
    decays more slowly than that of the global block. Bottom: Number of shots required for a gradient component to reach the shot-noise floor versus register width.}
    \label{fig:perf_noise}
  \end{minipage}

\end{figure*}



\subsubsection{Trainability Under Depolarizing Noise and Measurement Cost}
\label{sec:perf_noise}

On near-term quantum hardware, however, gate noise can further suppress gradients and produce a noise-induced barren plateau~\cite{wang2021noise}. We therefore compare how the Q-Capsule and global architectures respond to noise. A depolarizing channel with strength $p$ is applied after each circuit layer, and the mean gradient magnitude is measured as $p$ increases. Figure~\ref{fig:perf_noise} presents the results at $N=6$, where mixed-state simulation remains computationally feasible. The Q-Capsule gradients decrease more slowly than those of the global block and remain larger at every tested noise level. At $p=0$, the mean gradient magnitudes are $9.8\times10^{-2}$ for Q-Capsule and $5.4\times10^{-2}$ for the global block. At $p=0.1$, they decrease to $5.1\times10^{-2}$ and $1.5\times10^{-2}$, respectively. Thus, although noise weakens both models, its effect is substantially stronger on the global architecture. This result suggests that the shallow and local entangling structure of Q-Capsule reduces noise accumulation along the paths responsible for carrying gradient information. The bottom panel translates the noise-free gradient advantage into measurement cost. Because the number of shots required to resolve a gradient scales approximately as
$1/\operatorname{Var}[\partial_\theta\mathcal{L}]$, the measurement requirement of Q-Capsule remains nearly constant as the register width increases. In contrast, the required number of shots for the global block grows by several orders of magnitude. Therefore, the global gradients are not exactly zero, but resolving them against shot noise becomes increasingly impractical at larger register widths.

\begin{table}[htb!]
  \centering
  \caption{Digit classification by three depth policies of the Q-Capsule circuit
  ($N=6$, $c_{\text{size}}=3$, $k=8$). The adaptive policy's converged depth
  profile is shown in the last column. Best accuracy per task in bold; the
  adaptive policy also uses roughly a quarter of the gates and parameters of
  the fixed-deep circuit.}
  \label{tab:classification}
  \small
  \setlength{\tabcolsep}{4pt}
  \setlength{\aboverulesep}{0pt}
  \setlength{\belowrulesep}{0pt}
  \setlength{\extrarowheight}{2pt}
  \begin{tabular}{l
                  >{\columncolor{colteal}}l
                  >{\columncolor{colsteel}}c
                  >{\columncolor{colgreen}}c
                  >{\columncolor{colpeach}}c
                  >{\columncolor{colcream}}c
                  >{\columncolor{colmauve}}c}
    \toprule
    \textbf{Task}
      & \cellcolor{hdrteal}\textbf{Depth policy}
      & \cellcolor{hdrsteel}\textbf{Accuracy}
      & \cellcolor{hdrgreen}\textbf{2-qubit gates}
      & \cellcolor{hdrpeach}\textbf{Total gates}
      & \cellcolor{hdrcream}\textbf{Params}
      & \cellcolor{hdrmauve}\textbf{Depths} \\
    \midrule
      & Fixed-shallow $(d{=}1)$ & $0.491$          & $7$  & $25$  & $24$  & $[1,1]$ \\
      & Fixed-deep $(d{=}8)$    & $0.944$          & $49$ & $151$ & $192$ & $[8,8]$ \\
    \multirow{-3}{*}{Binary ($3$/$8$)}
      & \textbf{Adaptive}       & $\mathbf{0.981}$ & $13$ & $43$  & $48$  & $[2,2]$ \\
    \midrule
      & Fixed-shallow $(d{=}1)$ & $0.935$          & $7$  & $25$  & $24$  & $[1,1]$ \\
      & Fixed-deep $(d{=}8)$    & $0.963$          & $49$ & $151$ & $192$ & $[8,8]$ \\
    \multirow{-3}{*}{Multiclass ($0$--$3$)}
      & \textbf{Adaptive}       & $\mathbf{0.977}$ & $13$ & $43$  & $48$  & $[2,2]$ \\
    \bottomrule
  \end{tabular}
\end{table}

\subsection{Classification Performance}
\label{sec:classification}
We evaluate the fixed-shallow, fixed-deep, and adaptive Q-Capsule policies on binary digit classification ($3$/$8$) and four-class classification
($0$--$3$). The experiments use an $N=6$ register divided into two capsules of size $c_{\text{size}}=3$. The binary task is read out from the capsule-head qubits and thresholded at zero; the four-class task reads all $N$ Pauli-$Z$ expectations into a trained linear head. Table~\ref{tab:classification} reports the test accuracy, circuit cost, parameter count, and final capsule depths.

Table~\ref{tab:classification} reports the outcome. The two fixed policies bracket the useful regime from either side, and neither is a good default. The shallow policy at $d{=}1$ is not merely weak on the binary task but collapses to chance ($0.491$ on a balanced two-class test set), indicating that a single layer per capsule leaves the circuit unable to separate the two digits at all; on the four-class task the same policy reaches $0.935$, so the failure is specific to the binary setting rather than a general lack of capacity. The deep policy at $d{=}8$ is reliable on both tasks ($0.944$ and $0.963$) but pays for it, using $49$ two-qubit gates and $192$ parameters---roughly four times the gate budget and eight times the parameter count of the shallow policy---and it is the slowest variant to train on both tasks.

\begin{figure*}[htb!]

\begin{minipage}[htb!]{0.48\textwidth}
  \centering
  \includegraphics[width=\textwidth]{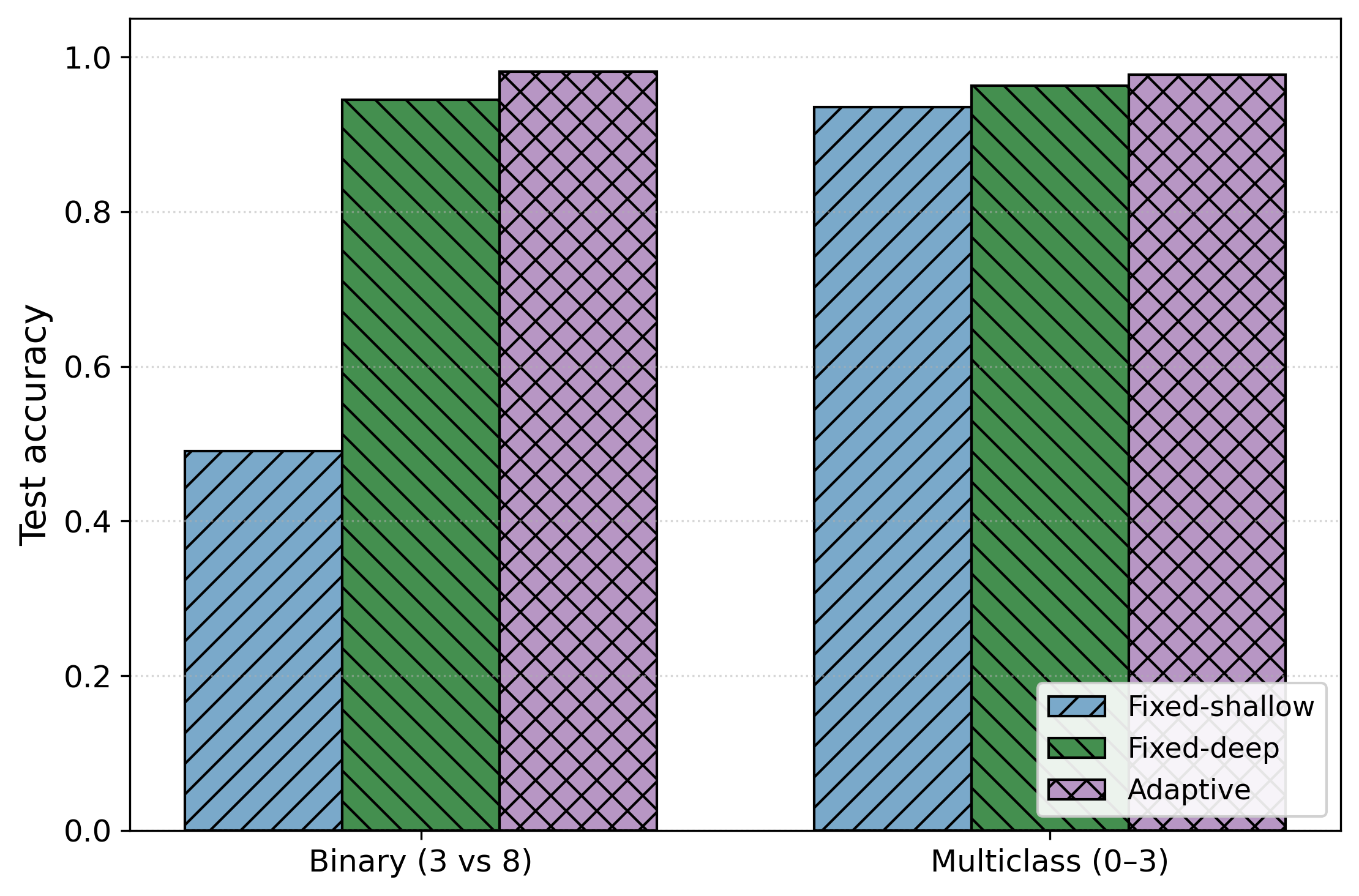}
  \caption{Test accuracy per task for the three depth policies. The shallow policy
  ($d{=}1$) collapses to chance on the binary task while remaining serviceable on the
  four-class task; the deep policy ($d{=}8$) is reliable on both but at four times the
  two-qubit gate count; the adaptive policy matches or exceeds the deep policy on both
  tasks using $13$ two-qubit gates rather than $49$.}
  \label{fig:classification}

\end{minipage}
\hfill
\begin{minipage}[htb!]{0.48\textwidth}
  
\centering
  \centering
  \captionof{table}{Cross-framework benchmark of the $20$-layer Q-Capsule circuit at
  capsule-aligned register widths.}
  \label{tab:fw_bench}
  \setlength{\aboverulesep}{0pt}
  \setlength{\belowrulesep}{0pt}
  \setlength{\extrarowheight}{2pt}
  \resizebox{\textwidth}{!}{%
  \begin{tabular}{l|
                  >{\columncolor{colsteel}}c
                  >{\columncolor{colcream}}c
                  >{\columncolor{colgreen}}c
                  >{\columncolor{colmauve}}c}
    \toprule
    & \cellcolor{hdrsteel}
    & \cellcolor{hdrcream}
    & \cellcolor{hdrgreen}\textbf{Gradient}
    & \cellcolor{hdrmauve}\textbf{Gradient} \\
    \multirow{-2}{*}{\textbf{Framework}}
    & \cellcolor{hdrsteel}\multirow{-2}{*}{\textbf{$N$}}
    & \cellcolor{hdrcream}\multirow{-2}{*}{\textbf{Params}}
    & \cellcolor{hdrgreen}\textbf{variance}
    & \cellcolor{hdrmauve}\textbf{time (s)} \\
    \midrule
    \multirow{5}{*}{PennyLane}
      & $6$  & $480$  & $1.892\times10^{-2}$ & $2.30\times10^{1}$ \\
      & $9$  & $720$  & $1.245\times10^{-2}$ & $4.39\times10^{1}$ \\
      & $12$ & $960$  & $1.350\times10^{-2}$ & $7.86\times10^{1}$ \\
      & $15$ & $1200$ & $1.021\times10^{-2}$ & $1.93\times10^{2}$ \\
      & $18$ & $1440$ & $1.252\times10^{-2}$ & $1.36\times10^{3}$ \\
    \midrule
    \multirow{5}{*}{Qiskit}
      & $6$  & $480$  & $1.892\times10^{-2}$ & $4.11\times10^{-2}$ \\
      & $9$  & $720$  & $1.245\times10^{-2}$ & $6.28\times10^{-2}$ \\
      & $12$ & $960$  & $1.350\times10^{-2}$ & $1.16\times10^{-1}$ \\
      & $15$ & $1200$ & $1.021\times10^{-2}$ & $5.49\times10^{-1}$ \\
      & $18$ & $1440$ & $1.252\times10^{-2}$ & $7.73\times10^{0}$ \\
    \midrule
    \multirow{5}{*}{Cirq}
      & $6$  & $480$  & $1.891\times10^{-2}$ & $6.33\times10^{-2}$ \\
      & $9$  & $720$  & $1.245\times10^{-2}$ & $9.43\times10^{-2}$ \\
      & $12$ & $960$  & $1.349\times10^{-2}$ & $1.42\times10^{-1}$ \\
      & $15$ & $1200$ & $1.021\times10^{-2}$ & $2.36\times10^{-1}$ \\
      & $18$ & $1440$ & $1.252\times10^{-2}$ & $1.50\times10^{0}$ \\
    \bottomrule
  \end{tabular}%
  }
\end{minipage}
\end{figure*}

The adaptive policy is the most accurate variant on both tasks, reaching $0.981$ on the binary task and $0.977$ on the four-class task, in each case at $13$ two-qubit gates and $48$ parameters. Relative to the deep policy this is a $73\%$ reduction in two-qubit gates and a $75\%$ reduction in parameters, obtained without a loss in accuracy. The mechanism is visible in the depth-selection trace: the probe assigns the binary task quick accuracies of $0.480$, $0.940$, and $0.960$ at $d=1,2,3$ and the four-class task $0.871$, $0.921$, and $0.911$, so in both cases $d{=}2$ is the shallowest depth within tolerance of the best probe, and both runs are initialized there. Neither run subsequently improved on the validation accuracy recorded at that depth, and both therefore retained the $[2,2]$ profile. The result is thus driven by the depth-selection stage rather than by subsequent growth: on these two tasks the policy's contribution is to identify a depth that neither fixed setting supplies, one layer above the collapsing shallow circuit and six below the deep one. We note that the accuracy margins over the deep policy are small in absolute terms---four test samples on the binary task ($n=108$) and three on the four-class task ($n=217$)---so the accuracy ordering between the adaptive and deep policies should be read as a tie, while the difference in circuit cost is large and unambiguous. Figure~\ref{fig:classification} shows the accuracy comparison; every variant except the shallow policy on the binary task sits well above the per-task chance level.

\subsection{Cross-Framework Implementation Benchmark}
\label{sec:framework}
To verify that the reported gradient behavior is a property of Q-Capsule rather than a specific software framework, we implemented the same circuit in PennyLane, Qiskit, and Cirq. PennyLane and Qiskit compute gradients using the exact parameter-shift rule, while Cirq uses a central finite-difference approximation with step size $\varepsilon=10^{-3}$. Table~\ref{tab:fw_bench} shows that all three implementations produce nearly identical gradient variances. PennyLane and Qiskit agree to numerical precision, while Cirq differs by less than approximately $0.05\%$. For example, at $N=6$, PennyLane and Qiskit obtain $1.8918\times10^{-2}$, compared with $1.8909\times10^{-2}$ for Cirq. Across all tested widths, the variance remains within an $\mathcal{O}(10^{-2})$ range and shows no systematic decay. This confirms that Q-Capsule's stable gradient behavior is independent of the simulator and differentiation method.

However, these frameworks differ more clearly in gradient-evaluation time, as shown in Table~\ref{tab:fw_bench}. At $N=6$, PennyLane requires $2.30\times10^{1}$~s to assemble the complete
gradient vector, whereas Qiskit and Cirq need only $4.11\times10^{-2}$~s and
$6.33\times10^{-2}$~s respectively for a single component. At $N=18$, PennyLane
requires approximately $1.36\times10^{3}$~s, compared with $7.73$~s for Qiskit
and $1.50$~s for Cirq. This difference mainly results from the gradient components being evaluated. PennyLane computes the complete gradient vector using parameter shift, requiring two circuit evaluations for each of the $4NL$ parameters. Qiskit and Cirq evaluate only the single gradient component needed for this experiment. Therefore, these results should not be interpreted as a direct comparison of simulator speed. Instead, they show that full-vector parameter-shift differentiation becomes expensive as the number of parameters increases. Circuit construction remains negligible in all frameworks, with build times below $6\times10^{-4}$~s.

\section{Conclusion}
\label{sec:conclusion}

This work introduced \textsc{Q-Capsule}, a localized quantum neural architecture that mitigates barren plateaus through capsule partitioning, local readout, sparse inter-capsule coupling, trainable data re-uploading, and QFIM-guided adaptive growth. Q-Capsule consistently maintains stable gradient variance,
demonstrating robust trainability across different register sizes. At $N=16$ and depth $25$, \textsc{Q-Capsule} achieves a gradient variance of $8.3\times10^{-2}$, compared with approximately $10^{-6}$ for the globally read baselines. Ablation studies further confirm the importance of capsule locality, sparse coupling, data re-uploading, and structured initialization, while the QFIM-based analysis shows that adaptive growth increases effective representational capacity.

Beyond gradient preservation, \textsc{Q-Capsule} provides structured optimization landscapes, higher parameter efficiency, improved robustness to depolarizing noise, and lower measurement requirements. Its adaptive policy achieves $0.981$ accuracy on binary classification and $0.977$ on four-class classification, while using about $73\%$ fewer two-qubit gates than the fixed-deep model on the multiclass task. Consistent gradient statistics across PennyLane, Qiskit, and Cirq further confirm that these improvements arise from the circuit design rather than a particular software framework. Overall, \textsc{Q-Capsule} provides a practical balance between trainability, expressivity, and quantum resource efficiency, motivating future evaluation on larger tasks and real noisy quantum hardware. Though \textsc{Q-Capsule} introduces additional training and computational overhead through periodic QFIM evaluations, data re-uploading, and adaptive circuit reconstruction, this overhead is primarily incurred during training and does not affect inference once the final circuit structure is fixed. Future work will focus on reducing this overhead, scaling the architecture to more complex datasets and larger quantum systems, and validating its performance on real noisy quantum processors.

\section{Acknowledgments}

The authors acknowledge the computational resources and support provided by Secure Prediction, Edge AI and Multimodal LLM Lab (SPEED Lab), Institute for Smarter Cities, Spaces, and Health, and Center for Connected Autonomy and Artificial Intelligence (CA-AI) at Florida Atlantic University. 

\section*{Ethics and Privacy Statement}

This work develops optimization techniques for variational quantum machine learning models with the goal of improving training stability, convergence, and scalability. The proposed methods are intended to advance trustworthy and efficient quantum AI research and do not involve human subjects, personal data, or sensitive datasets. While improved quantum learning algorithms may accelerate future applications across scientific and engineering domains, the methods presented are general-purpose optimization techniques and do not introduce application-specific privacy, fairness, or safety risks beyond those inherent to quantum machine learning research.

\bibliographystyle{plainnat}
\bibliography{sections/References}

\end{document}